\documentclass[11pt]{article}

\usepackage{color,graphicx}
\usepackage{young}
\usepackage[vcentermath]{youngtab}
\usepackage{amsmath,amssymb,graphicx}
\usepackage{hyperref}
\usepackage{xtab}
\definecolor{darkred}{rgb}{0.65,0.15,0}
\hypersetup{pdfborder={0 0 0},colorlinks=true,urlcolor=darkred,citecolor=blue,linkcolor=darkred,linktocpage=true}

\usepackage{cite}
\usepackage{amsmath}
\usepackage{amsfonts}
\usepackage{amssymb}
\usepackage{graphicx}%
\usepackage{amsthm}
\usepackage{mathrsfs}
\usepackage[T1]{fontenc}
\usepackage{enumerate}
\usepackage{color}

\def\4diml{four-dimensional}

\def\-1{^{-1}}

\newcommand{\M}{\mathscr{M}}

\makeatletter

\@addtoreset{equation}{section}
\makeatother

\begin{document}

\thispagestyle{empty}

\vspace{5mm}

\begin{center}
{\LARGE \bf Integrable deformed $H_{_{4}}$ WZW models and their non-Abelian duals as solutions of generalized supergravity equations}

\vspace{15mm}
\normalsize
{\large Ali Eghbali\footnote{eghbali978@gmail.com}, Simin Ghasemi-Sorkhabi\footnote{s.ghassemi.s@gmail.com}, Adel Rezaei-Aghdam\footnote{Corresponding author:rezaei-a@azaruniv.ac.ir}
}

\vspace{2mm}
{\small \em Department of Physics, Faculty of Basic Sciences,\\
Azarbaijan Shahid Madani University, 53714-161, Tabriz, Iran}\\

\vspace{7mm}


\vspace{6mm}

\begin{tabular}{p{12cm}}
{\small
We show that the Yang-Baxter (YB) deformed backgrounds of the Wess-Zumino-Witten (WZW) model based on the $H_{_{4}}$ Lie group
can be considered as solutions of the generalized supergravity equations (GSEs).
Then, by applying the Poisson-Lie T-duality in the presence of spectator fields, we obtain the non-Abelian target space duals of those models.
It is shown that all dual models (except for one model) are integrable
and most interestingly, they satisfy the GSEs.
As a final remark, we show that all solutions of the GSEs for the original models are trivial.
}
\end{tabular}
\vspace{-1mm}
\end{center}

{\hspace{1cm}Keywords:} Generalized supergravity,  $\sigma$-model, String duality, Wess-Zumino-\\
$~~~~~~~~~~~~~~~~~~~~~~~~~~~~~$ Witten model, Yang-Baxter deformation

\setcounter{page}{1}
\newpage
\tableofcontents

 \vspace{2mm}

\vspace{-1mm}
\section{\label{Sec.I} Introduction}
Over the last forty five years, significant interest has developed around integrable two-dimensi-
\\
onal $\sigma$-models and their deformations
\cite{Pohlmeyer,Zakharov,Eichenherr}.
The initial presentation of the integrable deformation of the principal chiral model on $SU(2)$ can be found in refs. \cite{Cherednik,Balog,Fateev,Sochen}.
Klimcik introduced the YB deformation of the chiral model \cite{Klimcik1,Klimcik2,Klimcik3}.
The YB $\sigma$-model is based on R-operators, which satisfy the (modified) classical YB equations \cite{Matsumoto,Kawaguchi,Kyono,Eghbali1,Eghbali2}.
The application of these integrable deformations to string theory specially the $AdS_5\times S^5$ string model
has been discussed  in \cite{delduc2014integrable,Kawaguchi,supercoset} (see, also, \cite{HidekiKyono,Arutyunov1,Hoare1}).
It has been shown that for homogeneous YB deformed models \cite{supergeometry} there is no Weyl anomaly if the $R$-operators are unimodular (see, also, \cite{Two-loop} up to two-loop, and \cite{Hronek}).
In \cite{Sheikh-Jabbari1}, the relationship between unimodularity condition on $R$-matrices with the divergence-free of the noncommutative parameter $\Theta$ of the dual noncommutative gauge theory has been mentioned;
moreover, it has been shown that the GSEs \cite{invariance} reproduce the classical YB equations, in such a way that $\Theta$ is the most general $r$-matrix solution
built from anti-symmetric products of Killing vectors \cite{Sheikh-Jabbari2}.
The $r$-matrices may be sorted into Abelian and non-Abelian types, and it has been proved that Abelian $r$-matrices are associated with T-duality shift T-duality transformations \cite{Tongeren},
thus ensuring that the corresponding YB deformation is a supergravity solution.
Regarding non-Abelian $r$-matrices, the unimodularity condition on the $r$-matrix \cite{supergeometry}  distinguishes valid
supergravity backgrounds \cite{callan1985strings} from the solutions of GSEs \cite{invariance}.
One of the differences between standard supergravity and GSEs is that the conventional dilaton is hidden
in one-form $Z$, which is one of the building blocks of the GSEs.
Artyunov and his colleagues introduced the GSEs to examine integrable deformations of the type $II$ superstring $\sigma$-model $AdS_5 \times S^5$ which is closely linked to non-Abelian T-duality transformations \cite{invariance}.
The GSEs in string theory include additional vector field $I$, compared to standard type $IIB$ supergravity, while the corresponding classical action is unknown, the equations of motion have been established.
The solutions of standard supergravity can be mapped to the solutions of the GSEs via T-duality, highlighting the equal importance of both in string theory \cite{generalized}.
Tseytlin and Wulff demonstrated that the GSEs can be derived by solving the $\kappa$-symmetry constraints of the Green-Schwarz action \cite{generalized}.
Then, in order to generalize the calculations of Tseytlin and Wulff,  it was constructed \cite{w.muck} a suitable counterterm (generalized
Fradkin-Tseytlin counterterm) for background fields satisfying the GSEs.

In recent years, we have witnessed further interest in the GSEs.
In \cite{Sakamoto1}, it was shown that the bosonic part of the GSEs can be completely reproduced from the modified double field theory \cite{{DFT1},{DFT2},{DFT3}}. Also,
it was shown that the equations of motion of the double field theory lead to the
GSEs when the dilaton has a linear dual-coordinate dependence (see, also, \cite{yuho2}).
Recently, the Weyl invariance of bosonic string theories in generalized supergravity backgrounds has been demonstrated at
one-loop order by construction of a local counterterm \cite{yuho3}.
In \cite{M.Y.E.Colgain}, it has been shown that the non-Abelian target space duals corresponding to some of the Bianchi cosmologies are indeed solutions of the GSEs.
Recently, a formula for the vector field $I$ together with a transformation rule for dilaton field have been obtained by applying Poisson-Lie T-plurality \cite{vonUnge} on Bianchi cosmologies \cite{hlavaty1}.
Then, it has been shown that plural backgrounds together with the introduced dilaton field and $I$ satisfy the GSEs (see, also, \cite{{hlavaty2},{hlavaty3}}).
In our previous work \cite{Eghbali3} we showed that the BTZ metric \cite{Banadoc,Horowitz} is a solution of the GSEs.
Also in \cite{Eghbali4} we showed that the backgrounds of WZW models on the Lie groups $SL(2,\mathbb R)$, $GL(2,\mathbb R)$, $H_{_{4}}$,
$A_{_{4,10}}$ and $A_{_{5,3}}$ \cite{Patera} are solutions of the GSEs.
In the present work, we show that the YB deformed backgrounds of $H_{_{4}}$ \cite{Eghbali1} can be considered as solutions of the GSEs.
In addition, using the Poisson-Lie T-duality approach in the presence of spectator fields, we examine the non-Abelian T-dualization of the deformed models
and then show that all dual models (except for one model) are integrable and most interestingly, they satisfy the GSEs.
It's worth mentioning that the non-Abelian T-duality of the YB deformed WZW models on $GL(2,\mathbb{R})$ Lie group has recently been performed
in \cite{Eghbali2}. There, it has been shown that the deformed models
can be obtained as original models of Poisson-Lie T-dual $\sigma$-models constructed on a $2+2$-dimensional manifold
${\M}$ with the two-dimensional non-Abelian Lie group and its Abelian dual pair.

The plan of paper is as follows:
In section \ref{Sec.II}, we present a brief summary of the GSEs, where important formulas are outlined.
Then, it is shown that the backgrounds of YB deformations of the $H_{_{4}}$ WZW model constructed in \cite{Eghbali1} (Table 1) are the solutions of the GSEs, the results are summarized in Table 2.
As in \cite{Eghbali2} in section \ref{Sec.III} we will show that the YB deformed models are fundamentally equivalent to non-Abelian T-dual $\sigma$-models;
moreover, in this section we present the spectator-dependent background matrices for all the deformed models in Table 3.
At the end of section \ref{Sec.III}, we obtain the non-Abelian T-dual spaces of the deformed models.
In section \ref{Sec.IV}, by using the method given in \cite{Mohammedi},
we discuss the integrability of the non-Abelian T-dual $\sigma$-models corresponding to the deformed backgrounds of $H_{_{4}}$ WZW model.
In this way, the corresponding lax pairs are presented in Table 5.
In section \ref{Sec.V}, we examine that the non-Abelian T-dual backgrounds of the deformed $H_{_{4}}$ WZW models are also solutions of the GSEs;
the results are presented in Table 6.
At the end of this section, we discuss the triviality of solutions of the GSEs of both original and dual models, in such a way that the results are summarized in Table 7.
The last section is devoted to the final discussion of the results.
\vspace{-3mm}
\section{\label{Sec.II}  The YB deformed $H_{_{4}}$ WZW models as solutions of the GSEs}
In this section, we first give a brief review of the GSEs.
Then, we show that the YB deformed backgrounds of the $H_{_{4}}$ WZW model, which have been classified into ten distinct classes in ref. \cite{Eghbali1},
are solutions of the GSEs. The deformed backgrounds including metric and B-field are presented in Table 1.
As an example, we will also discuss in details the case of $H^{^{(\kappa, \eta)}}_{_{4}}.{II}$.

\subsection{A short review of the GSEs}\label{Sec.II.1}
The bosonic GSEs in D dimensions in the absence of the Ramond-Ramond fields take the following form \cite{Arutyunov1}:
\begin{eqnarray}
{\cal R}_{_{MN}}-\frac{1}{4} H_{_{MPQ}} {H^{^{PQ}}}_{_{N}}+\nabla_{_{M}} X_{_{N}} +{\nabla}_{_{N}} X_{_{M}}  &=& 0,\label{2.1}\\
\frac{1}{2}{\nabla}^{^{R}} H_{_{RMN}} -X^{^{R}} H_{_{RMN}}-{\nabla}_{_{M}} X_{_{N}} +{\nabla}_{_{N}} X_{_{M}} &=&0,\label{2.2}\\
{\cal R}-\frac{1}{12}  H^{^{2}} +4 {\nabla}_{_{M}} X^{^{M}} -4 X_{_{M}} X^{^{M}}-4 \Lambda &=& 0,\label{2.3}
\end{eqnarray}
where ${\cal R}_{_{MN}}$ and ${\cal R}$ are the respective Ricci tensor and scalar curvature of the metric $G_{_{MN}}$, and $\Lambda$ is the cosmological constant.
Here, the $D$-dimensional indices $M, N,...$ of coordinates $x^{M}$ of manifold $\M$ are raised or lowered with the metric $G_{_{MN}}.$
The covariant derivative ${\nabla}_{_{M}}$ is the conventional Levi-Civita connection associated to $G_{_{MN}}.$
The $X_{_M}$ are related to the one-form $Z$ via $X_{_{M}}= I_{_{M}}+ Z_{_{M}}$, such that the vector field $I=I^{^M} \partial_{_{M}}$  and a one-form $Z=Z_{_{M}}dx^{^{M}}$ are defined so as to satisfy
\begin{eqnarray}
{\cal L}_{_{I}} G_{_{MN}}&=&0,\label{2.4}\\
{\cal L}_{_{I}} B_{_{MN}}&=&0,\label{2.5}\\
{\nabla}_{_{M}} Z_{_{N}} -{\nabla}_{_{N}} Z_{_{M}} +I^{^{R}} H_{_{RMN}}  &=&0,\label{2.6}\\
I^{^{M}} Z_{_{M}}&=&0,\label{2.7}
\end{eqnarray}
where ${\cal L}$ stands for the Lie derivative.


\begin{center}
\small {{{\bf Table 1.}~ YB deformed backgrounds of the $H_{_{4}}$ WZW model \cite{Eghbali1} and their Killing vectors}}
{\scriptsize
\renewcommand{\arraystretch}{1.2}{
\begin{tabular}{lll} \hline \hline
Background symbol &  Backgrounds including metric and $B$-field  &  Killing vectors\\
\hline
				          &                                            & $K_{_{1}}=-\partial_{_{x}}+u \partial_{_{u}}-\rho \partial_{_{v}},$\\
                          &                                            & $K_{_{2}}=e^{-x}\partial_{_{y}}-u \partial_{_{v}},$                \\
{$H^{(\kappa)}_{_{4}}.I$} & $ds^{2}=\rho dx^{2}-2dx dv+ 2e^{x} dy du,$ & $K_{_{3}}=\partial_{_{y}},$                                        \\
                          &\hspace{1mm} $B=\kappa y e^{x} du\wedge dx$ & $K_{_{4}}=y\partial_{_{y}}-u\partial_{_{u}},$                      \\
                          &                                            & $K_{_{5}}=e^{-x}\partial_{_{u}}-y\partial_{_{v}},$                 \\
                          &                                            & $K_{_{6}}=\partial_{_{u}},$                                        \\
                          &                                            & $K_{_{7}}=-\partial_{_{v}}.$                                       \\
\hline
				               &                                                           & $K_{_{1}}=-\partial_{_{x}}+u\partial_{_{u}}+(2\eta^2-\rho)~\partial_{_{v}},$  \\
                               &                                                           & $K_{_{2}}=e^{-x}\partial_{_{y}}-u \partial_{_{v}},$                           \\
$H^{(\kappa, \eta)}_{_{4}}.II$ & $ds^{2}= (\rho-2\eta^{2})dx^{2}-2  dx dv+ 2 e^{x} dy du,$ & $K_{_{3}}=\partial_{_{y}},$                                                   \\
                               &\hspace{1mm} $B=\kappa y e^{x} du\wedge dx$                & $K_{_{4}}=y\partial_{_{y}}-u\partial_{_{u}},$                                 \\
                               &                                                           & $K_{_{5}}=e^{-x}\partial_{_{u}}-y\partial_{_{v}},$                            \\
                               &                                                           & $K_{_{6}}=\partial_{_{u}},$                                                   \\
                               &                                                           & $K_{_{7}}=-\partial_{_{v}}.$                                                  \\
\hline                              	
				                          &                                                                       &$K_{_{1}}=-\partial_{_{x}}+u \partial_{_{u}}-\rho \partial_{_{v}},$                          \\
                                          &                                                                       &$K_{_{2}}=e^{-x}\partial_{_{y}}-u \partial_{_{v}},$                                          \\
$H^{(\kappa,\eta,\tilde{A})}_{_{4}}.{III}$& $ds^{2}=\rho dx^{2}-2 dx dv+2 e^{x} dy du-\rho \eta^{2}e^{2x}du^{2},$ &$K_{_{3}}=-\frac{1}{\rho \eta^2}\partial_{_{u}},$                                            \\
                                          &\hspace{1mm} $B=\kappa y e^{x} du\wedge dx+\tilde{A} e^{x}dv\wedge du$ &$K_{_{4}}=\partial_{_{y}},$                                                                  \\
                                          &                                                                       &$K_{_{5}}=(x-1)\partial_{_{y}}$                                                              \\
                                          &                                                                       &\hspace{5mm}$-\frac{e^{-x}}{\rho\eta^2}~\partial_{_{u}}+\frac{y}{\rho\eta^2}~\partial_{_{v}},$\\
                                          &                                                                       &$K_{_{6}}=-\partial_{_{v}}.$                                                                 \\
\hline
				                          &                                                                            & $K_{_{1}}=(\eta^2-1)\partial_{_{x}}-y\eta^2~\partial_{_{y}}$                                      \\
                                          &                                                                            & \hspace{5mm}$+u\partial_{_{u}}+\rho(\eta^2-1)~\partial_{_{v}},$                                   \\
                                          &                                                                            & $K_{_{2}}=e^{-x}\partial_{_{y}}-u\partial_{_{v}},$                                                \\
$H^{(\kappa, \eta, \tilde{A})}_{_{4}}.IV$ &$ds^{2}=\frac{1}{1-\eta^{2}}\big[\rho  dx^{2}-2 dx dv$                      & $K_{_{3}}=e^{-\frac{x\eta^2}{(\eta^2-1)}}\partial_{_{y}},$                                        \\
                                          &\hspace{5mm}$-2\eta^{2}  y e^{x} dx du\big] +2 e^{x} dy du,$                & $K_{_{4}}=y\partial_{_{y}}-u\partial_{_{u}},$                                                     \\
                                          &\hspace{1.5mm}$B=(\kappa- \frac{\tilde{A}}{1-\eta^{2}}) ye^{x} du\wedge dx$ & $K_{_{5}}=\partial_{_{u}},$                                                                       \\
                                          &                                                                            & $K_{_{6}}=e^{\frac{x}{(\eta^2-1)}}\partial_{_{u}}-e^{\frac{ x\eta^2}{(\eta^2-1)}}y\partial_{_{v}},$\\                                      &                                                                            & $K_{_{7}}=(\eta^2-1)\partial_{_{v}}.$                                                             \\
\hline
				                        &                                                            & $K_{_{1}}=-\partial_{_{x}}+\frac{2\eta^2}{\eta^2-1}\partial_{_{y}}+u\partial_{_{u}}-\rho\partial_{_{v}},$\\
                                        & $ds^{2}=\rho  dx^{2}- 2 dx dv+ 2 e^{x} dy du $             & $K_{_{2}}=\partial_{_{y}},$                                                        \\
$H^{(\kappa, \eta, \tilde{A})}_{_{4}}.V$&\hspace{5.3mm}$-\frac{4 \eta^{2}}{1-\eta^{2}} e^{x} dx du,$ & $K_{_{3}}=e^{-x}\partial_{_{y}}-u\partial_{_{v}},$                                 \\
                                        &\hspace{1.9mm}$B=(\kappa + \tilde{A}) y e^{x} du\wedge dx$  & $K_{_{4}}=\frac{\eta^2(y-2+2x)-y}{\eta^2-1}\partial_{_{y}}-u\partial_{_{u}},$      \\
                                        &                                                            & $K_{_{5}}=e^{-x}\partial_{_{u}}-\frac{\eta^2(y-2+2x)-y}{\eta^2-1}\partial_{_{v}},$ \\
                                        &                                                            & $K_{_{6}}=\partial_{_{u}},$                                                        \\
                                        &                                                            & $K_{_{7}}=-\partial_{_{v}}.$                                                       \\
\hline
				                          &                                                                                   & $K_{_{1}}=-\partial_{_{x}}+\frac{\rho\eta^2}{\eta^2-1}~\partial_{_{y}}$       \\
                                          &                                                                                   &\hspace{5mm}$+u\partial_{_{u}}-\rho~\partial_{_{v}},$                          \\
                                          &$ds^{2}=\rho  dx^{2}-2 dx dv+2 e^{x} dy du$                                        & $K_{_{2}}=e^{-x}\partial_{_{y}}-u\partial_{_{v}},$                            \\
$H^{(\kappa, \eta, \tilde{A})}_{_{4}}.VI$ &\hspace{5.3mm}$-\frac{2\rho \eta^{2}}{1-\eta^{2}}e^{x} dx du-\frac{\rho \eta^{2}}{1-\eta^{2}}e^{2x}du^{2},$ & $K_{_{3}}=\frac{(\eta^2-1)}{\rho \eta^2}\partial_{_{u}},$ \\
                                          &\hspace{1.9mm}$B=(\kappa + \tilde{A}) y e^{x} du\wedge dx  +\tilde{A}  e^{x} dv \wedge du$          & $K_{_{4}}=\partial_{_{y}},$                                       \\
                                          &                                                                                   & $K_{_{5}}=(x-1)\partial_{_{y}}+\frac{e^{-x}(\eta^2-1)}{\rho \eta^2}\partial_{_{u}}$\\
                                          &                                                                                   &\hspace{5mm}$-\frac{-\rho\eta^2-y+y\eta^2+x\rho\eta^2}{\rho\eta^2}\partial_{_{v}},$ \\
                                          &                                                                                   & $K_{_{6}}=-\partial_{_{v}}.$                                                  \\
\hline
\hline
\end{tabular}}}
\end{center}

\begin{center}
\small {{{\bf Table 1.}~Continued.}}
{\scriptsize
\renewcommand{\arraystretch}{1.2}{
\begin{tabular}{lll} \hline \hline
Background symbol &  Backgrounds including metric and $B$-field  &  Killing vectors\\
\hline
				                          &                                                                                   &$K_{_{1}}=-\partial_{_{x}}+u\partial_{_{u}}-\frac{\rho}{\eta^2+1}~\partial_{_{v}},$ \\
                                          &                                                                                   &$K_{_{2}}=e^{-x}\partial_{_{y}}-u\partial_{_{v}},$                                  \\
                                          & $ds^{2}=\frac{\rho}{1+\eta^{2}} dx^{2}-2 dx dv$                                   &$K_{_{3}}=-\frac{\eta^2+1}{\rho \eta^2}\partial_{_{u}},$                            \\
$H^{(\kappa, \eta, \tilde{A})}_{_{4}}.VII$& \hspace{5.3mm}$+2 e^{x} dy du-\frac{\rho \eta^{2}}{1+\eta^{2}}   e^{2x}  du^{2},$ &$K_{_{4}}=\partial_{_{y}},$                                                         \\
                                          & \hspace{1.9mm}$B= \kappa y e^{x} du\wedge dx + \tilde{A}e^{x}  dv\wedge du$       &$K_{_{5}}=(x-1)\partial_{_{y}}-\frac{e^{-x}(\eta^2+1)}{\rho \eta^2}\partial_{_{u}}$ \\
                                          &                                                                                   &\hspace{5mm}$+\frac{y(\eta^2+1)}{\rho\eta^2}\partial_{_{v}},$                       \\
                                          &                                                                                   &$K_{_{6}}=-\partial_{_{v}}.$                                                        \\
\hline
				                 &                                                          & $K_{_{1}}=\frac{1}{\eta^2-1}(\partial_{_{x}}-y\eta^2~\partial_{_{y}})$ \\
                                 &                                                          &\hspace{5mm}$+u\partial_{_{u}}+\frac{\rho}{\eta^2+1}\partial_{_{v}},$                                                            \\
                                 &                                                          & $K_{_{2}}=e^{-x}\partial_{_{y}}-u\partial_{_{v}},$                                         \\
$H^{(\kappa, \eta)}_{_{4}}.VIII$ & $ds^{2}= (1-\eta^{2})(\rho dx^{2}-2 dx dv)$              & $K_{_{3}}=e^{-x\eta^2}\partial_{_{y}},$                                                    \\
                                 &\hspace{5.3mm}$+2 e^{x} dy du+ 2 \eta^{2} y e^{x} dx du,$ & $K_{_{4}}=y\partial_{_{y}}-u\partial_{_{u}},$                                              \\
                                 &\hspace{1.9mm}$B=\kappa y e^{x} du\wedge dx$              & $K_{_{5}}=\partial_{_{u}},$                                                                \\
                                 &                                                          & $K_{_{6}}=e^{x(\eta^2-1)}\partial_{_{u}}-ye^{x\eta^2}\partial_{_{v}},$                     \\
                                 &                                                          & $K_{_{7}}=\frac{1}{\eta^2-1}\partial_{_{v}}.$                                              \\
\hline
				                            &                                                                                               &$K_{_{1}}=-\frac{(\eta^2 q^4-1)}{\eta^2-1}(\partial_{_{x}}
                                                                                                                                             +\rho   \partial_{_{v}}) $                                               \\
                                            &                                                                                               &\hspace{5mm}$+\frac{y\eta^2(q^4-1)}{\eta^2-1}\partial_{_{y}}+u\partial_{_{u}},$ \\
$H^{(\kappa, \eta, \tilde{A})}_{_{4,q}}.IX$ &$ds^{2}=\frac{1-\eta^{2}}{1-\eta^{2}q^{4}}(\rho dx^{2}-2 dx dv)$                               &$K_{_{2}}=e^{-x}\partial_{_{y}}-u\partial_{_{v}},$                       \\
                                            &\hspace{5.3mm}$+2 e^{x} dy du+  \frac{2\eta^{2}(1-q^{4})}{1-\eta^{2}q^{4}} y e^{x} dx du,$     &$K_{_{3}}=e^{\frac{-x(q^4-1)\eta^2}{(\eta^2 q^4-1)}}\partial_{_{y}},$    \\
                                            &\hspace{1.9mm}$B=\Big[\kappa-\frac{\tilde{A}  q^{2} (1-\eta^{2})}{1-\eta^{2}q^{4}}\Big] y e^{x} du\wedge dx$ & $K_{_{4}}=y\partial_{_{y}}-u\partial_{_{u}},$             \\
                                            &                                                                                               &$K_{_{5}}=\partial_{_{u}},$                                              \\
                                            &                                                & $K_{_{6}}=e^{\frac{-x(\eta^2-1)}{(\eta^2q^4-1)}}\partial_{_{u}}-ye^{\frac{x(q^4-1)\eta^2}{(\eta^2q^4-1)}}\partial_{_{v}},$  \\
                                            &                                                                                               & $K_{_{7}}=-\frac{(\eta^2q^4-1)}{\eta^2-1}\partial_{_{v}}.$              \\
\hline
				                       &                                                           & $K_{_{1}}=-\partial_{_{x}}+u~\partial_{_{u}}-\rho~\partial_{_{v}},$ \\
                                       &                                                           & $K_{_{2}}=e^{-x}\partial_{_{y}}-u \partial_{_{v}},$                 \\
$H^{(\kappa, \tilde{A})}_{_{4}}.{X}$   & $ds^{2}=\rho  dx^{2}-2 dx dv+2 e^{x} dy du,$              & $K_{_{3}}=\partial_{_{y}},$                                         \\
                                       &\hspace{1.9mm}$B=(\kappa - \tilde{A}) y e^{x} du\wedge dx$ & $K_{_{4}}=y\partial_{_{y}}-u\partial_{_{u}},$                       \\
                                       &                                                           & $K_{_{5}}=e^{-x}\partial_{_{u}}-y\partial_{_{v}},$                  \\
                                       &                                                           & $K_{_{6}}=\partial_{_{u}},$                                          \\
                                       &                                                           & $K_{_{7}}=-\partial_{_{v}}.$                                        \\
\hline
\hline
\end{tabular}}}
\end{center}
The field strength $H_{_{MNP}}$ corresponding to anti-symmetric tensor field $B_{_{MN}}$ (B-field) is defined as
\begin{eqnarray}\label{2.8}
H_{_{MNP}}= \partial_{_{M}} B_{_{NP}}+\partial_{_{N}} B_{_{PM}}+\partial_{_{P}} B_{_{MN}}.
\end{eqnarray}
Note that the conventional dilaton is included in $Z_{_{M}}$  as follows:
\begin{eqnarray}\label{2.9}
Z_{_{M}}= \partial_{_{M}} \Phi + B_{_{NM}} I^{^{N}},
\end{eqnarray}
where  $\Phi$ is a scalar dilaton field.
Of course, one may use \eqref{2.9} to rewrite \eqref{2.7} in the form of $I^{^{M}}~ \partial_{_{M}} \Phi$=0.
A remarkable point is that if one sets  $I^{^{M}}$=0, then it is concluded that $X_{_{M}}=\partial_{_{M}} \Phi$
and thus, the GSEs turn into the standard supergravity equations.
In what follows, we shall show that the YB deformed backgrounds of the $H_{_{4}}$
WZW model can be considered as solutions of the GSEs (equations \eqref{2.1}-\eqref{2.7}).

\subsection{\label{Sec.II.2} An example: the $H^{^{(\kappa, \eta)}}_{_{4}}.{II}$ background as a solution for the GSEs}
Here, we investigate the YB deformed $H^{^{(\kappa, \eta)}}_{_{4}}.{II}$ background can be considered as a solution for the GSEs.
According to Table 1, the background of model including the metric and $B$-field is given by\footnote{It  seems  to  be  of  interest to define the line element and $B$-field corresponding to $G_{_{MN}}$ and $B_{_{MN}}$ in the coordinate basis.
They are, respectively, read
$$ds^2 = G_{_{MN}}d\phi^{^M}~d\phi^{^N},~~~ B= \frac{ 1}{2} ~B_{_{MN}} ~ d\phi^{^M} \wedge d\phi^{^N}.$$}
\begin{eqnarray}
ds^{2}&=&(\rho-2\eta^2) dx^{2}-2 dx dv+2 e^{x} dy du,\label{2.10}\\
B&=&\kappa y e^{x} du\wedge dx,\label{2.11}
\end{eqnarray}
where $(x,y,u,v)$ are coordinates on the Lie group $H_{_{4}}$, $\rho$ and $\kappa$ are some constant real parameters, while $\eta$ is a deformation parameter.
In the next step and to complete the calculations, the Killing vectors of the metric \eqref{2.10} are required.
They can be obtained by solving Killing equations, ${\cal L}_{_{K_a}}G_{_{MN}}=0$.
The Killing vectors of metric \eqref{2.10} are given as follows:
\begin{eqnarray}
K_{_{1}}&=&-\partial_{_{x}}+u~\partial_{_{u}}+(2\eta^2-\rho)~\partial_{_{v}},\nonumber\\
K_{_{2}}&=&e^{-x}\partial_{_{y}}-u~\partial_{_{v}},\nonumber\\
K_{_{3}}&=&\partial_{_{y}},\nonumber\\
K_{_{4}}&=&y\partial_{_{y}}-u\partial_{_{u}},\nonumber\\
K_{_{5}}&=&e^{-x}\partial_{_{u}}-y\partial_{_{v}},\nonumber\\
K_{_{6}}&=&\partial_{_{u}},\nonumber\\
K_{_{7}}&=&-\partial_{_{v}}.\label{2.12}
\end{eqnarray}
Now, we apply Killing vectors \eqref{2.12} to construct an appropriate vector field $I$.
To this end, one can use the following expansion
\begin{equation}\label{2.12.1}
I=\sum_{i=1}^{n}\alpha_{i}K_{i},
\end{equation}
where $n$ stands for the number of Killing vectors of the metric, and $\alpha_{_{i}}$'s are some constant parameters.
Applying this linear combination in equations \eqref{2.1}-\eqref{2.7} together with
metric \eqref{2.10} and $B$-field \eqref{2.11} we find two class of the solutions as follows:
\\
\begin{center}
\small {{{\bf Table 2.}~The YB deformed $H_{_{4}}$ WZW models as solutions of the GSEs}}
{\scriptsize
\renewcommand{\arraystretch}{1.6}{
\begin{tabular}{llll} \hline \hline
$H_{_{4}}$ &  Vector field~I  &  One-form Z & Comments \\
\hline
                        &$-\alpha_{_{1}} \partial_{_{x}}+\alpha_{_{4}} y \partial_{_{y}}$      & $[\alpha_{_{7}}-\frac{2\alpha_{_{4}}\alpha_{_{7}}}{\alpha_{_{1}}}+\alpha_{_{1}}\rho$
                        &$\Lambda=-8\alpha_{_{4}}[2\alpha_{_{7}}$\\
                        &$+[(\alpha_{_{1}}-\alpha_{_{4}})u$                                    & $-2\alpha_{_{4}}\rho +\alpha_{_{5}} y]~dx$
                        &~~~~$+(\alpha_{_{1}}-\alpha_{_{4}})\rho$\\
                        &$+\alpha_{_{5}}~e^{-x}+\alpha_{_{6}}]~\partial_{_{u}}$                & $+[e^{x}(-\alpha_{_{1}}u+\alpha_{_{4}}u-\alpha_{_{6}})+\alpha_{5}]dy$
                        &~~~~$-\frac{2\alpha_{_{4}}\alpha_{_{7}}}{\alpha_{_{1}}}],$\\
$H^{(\kappa)}_{_{4}}.I$ &$-(\alpha_{_{1}}~\rho+\alpha_{_{5}}~y+\alpha_{_{7}})~\partial_{_{v}}$ & $+y~e^x~\alpha_{_{4}}~du+(-\alpha_{_{1}}+2\alpha_{_{4}})dv$          & $\kappa=1$\\
\cline{2-4}
                        & $-\alpha_{_{7}} ~\partial_{_{v}}$   & $[\frac{x}{4}(1-\kappa^2)+c_{_{2}}]dx $ &  $\Lambda=0$ \\
\hline
&$-\alpha_{_{1}} \partial_{_{x}}+\alpha_{_{4}} y \partial_{_{y}}$      & $[\alpha_{_{7}}-\frac{2\alpha_{_{4}}\alpha_{_{7}}}{\alpha_{_{1}}}+\alpha_{_{5}} y$
                        &$\Lambda=-8\alpha_{_{4}}[2\alpha_{_{7}}$\\
                        &$+[(\alpha_{_{1}}-\alpha_{_{4}})u$                                    & $+(\alpha_{_{1}}-2\alpha_{_{4}})(\rho-2\eta^2) ]~dx$
                        &~~~~$+(\alpha_{_{1}}-\alpha_{_{4}})(\rho-2\eta^2)$\\
                        &$+\alpha_{_{5}}~e^{-x}+\alpha_{_{6}}]~\partial_{_{u}}$                & $+[e^{x}(-\alpha_{_{1}}u+\alpha_{_{4}}u-\alpha_{_{6}})+\alpha_{5}]dy$
                        &~~~~$-\frac{2\alpha_{_{4}}\alpha_{_{7}}}{\alpha_{_{1}}}],$\\
$H^{(\kappa,\eta)}_{_{4}}.II$ &$[\alpha_{_{1}}(\rho-2\eta^2)-\alpha_{_{5}}~y-\alpha_{_{7}}]~\partial_{_{v}}$ & $+y~e^x~\alpha_{_{4}}~du+(-\alpha_{_{1}}+2\alpha_{_{4}})dv$          & $\kappa=1$\\
\cline{2-4}
                        & $-\alpha_{_{7}} ~\partial_{_{v}}$   & $[\frac{x}{4}(1-\kappa^2)+c_{_{2}}]dx $ &  $\Lambda=0$ \\
\hline

$H^{(\kappa, \eta, \tilde{A})}_{_{4}}.{III}$ & $-\alpha_{_{6}}~\partial_{_{v}}$ & $[\frac{x}{4}(1-\kappa^2)+c_{_{2}}]dx$ &  $\Lambda=\tilde{A}=0$\\
\hline
                                           &                                              & $\{-\frac{x}{4(\eta^2-1)^2}[-1$               &\\
 $H^{(\kappa, \eta, \tilde{A})}_{_{4}}.IV$ & {$\alpha_{_{7}}~(\eta^2-1)~\partial_{_{v}}$} & $+\tilde{A}^2+\kappa^2(\eta^2-1)^2$           &$\Lambda=0$\\
                                           &                                              & $+2~\tilde{A}~\kappa(\eta^2-1)]+c_{_{2}}\}dx$ &\\
\hline
$H^{(\kappa, \eta, \tilde{A})}_{_{4}}.V$ & $-\alpha_{_{7}}~\partial_{_{v}}$                          & $\{\frac{x}{4}[1-(\kappa+\tilde{A})^2]+c_{_{2}}\}dx$            & $\Lambda=0$\\
\hline
$H^{(\kappa, \eta, \tilde{A})}_{_{4}}.VI$  & $-\alpha_{_{6}}~\partial_{_{v}}$  & $[\frac{x}{4}(1-\kappa^2)+c_{_{2}}]dx$  &  $\Lambda=\tilde{A}=0$\\
\hline
$H^{(\kappa, \eta, \tilde{A})}_{_{4}}.VII$ & $-\alpha_{_{6}}~\partial_{_{v}}$  & $[\frac{x}{4}(1-\kappa^2)+c_{_{2}}]dx$  &  $\Lambda=\tilde{A}=0$\\
\hline
                                 & $\frac{\alpha_{_{7}}}{\eta ^2-1}~\partial_{_{v}}$ & $[\frac{x}{4}(-\kappa^2+(\eta^2-1)^2+c_{_{2}}]dx$ & $\Lambda=0$\\
\cline{2-4}
                                 & $\frac{\alpha_{_{1}}}{\eta ^2-1}~\partial_{_{x}}$                        & $[\alpha_{_{6}}e^{x\eta^2}y$         & $\Lambda=\frac{1}{\alpha_{_{1}}(\eta^4-1)}\{8\alpha_{_{4}}(\alpha_{_{1}}-\alpha_{_{4}})$\\
                                 &$+[\alpha_{_{4}}y-\frac{\alpha_{_{1}}\eta^2 y}{\eta^2-1}]\partial_{_{y}}$ &$-y e^x \eta^2(u(\alpha_{_{1}}-\alpha_{_{4}})+\alpha_{_{5}})$ & $\times[\alpha_{_{7}}(1+\eta^2)$\\
$H^{(\kappa, \eta)}_{_{4}}.VIII$ &$[(\alpha_{_{1}}-\alpha_{_{4}})u+\alpha_{_{5}}$                           & $\frac{(\alpha_{_{1}}-2\alpha_{_{4}})(\alpha_{_{7}}(1+\eta^2)
                                 +\alpha_{_{1}}(\eta^2-1)\rho)}{\alpha_{_{1}}(1+\eta^2)}]dx$            & $+\alpha_{_{1}}\rho(\eta^2-1)]\},$\\
                                 &$+\alpha_{_{6}}e^{x(\eta^2-1)}]\partial_{_{u}}$      & $[-e^x\big(u(\alpha_{_{1}}-\alpha_{_{4}})+\alpha_{_{5}}\big)$      & $\kappa=1-\eta^2$ \\
                                 &$[\frac{\alpha_{_{1}}\rho}{\eta^2+1}-\alpha_{_{6}}ye^{x\eta^2}$
                                 &$+\alpha_{_{6}}e^{x\eta^2}]dy+\alpha_{_{4}}e^x~y~du$         & \\
                                 &$+\frac{\alpha_{_{7}}}{\eta ^2-1}]\partial_{_{v}}$ &$+(-\alpha_{_{1}}+2\alpha_{_{4}})dv$ &\\
\hline
                                            &                                                                     &$\Big\{\frac{-x}{4(-1+q^4\eta^2)^2}[\tilde{A}^2 q^4(\eta^2-1)^2$  &  \\
$H^{(\kappa, \eta, \tilde{A})}_{_{4,q}}.IX$ & {$-\frac{\alpha_{_{7}}(-1+\eta^2~q^4)}{\eta ^2-1}~\partial_{_{v}}$} &$-2\tilde{A}~\kappa ~q^2(\eta^2-1)(-1+q^4~\eta^2) $               & $\Lambda=0$\\
                                            &                                                                     &$+\kappa^2(-1+q^4~\eta^2)^2$                                      &\\
                                            &                                                                     &$-(1+(-2+q^4)\eta^2)^2]+c_{_{2}}\Big\}dx$                         &\\
\hline
                                  & $-\alpha_{_{7}} \partial_{_{v}}$ & $[\frac{x}{4}(1-(\tilde{A}-\kappa)^2)+c_{_{2}}]dx$  &  $\Lambda=0$\\
\cline{2-4}
                                  &$-\alpha_{_{1}} \partial_{_{x}}+\alpha_{_{4}} y \partial_{_{y}}$      & $[\alpha_{_{7}}-\frac{2\alpha_{_{4}}\alpha_{_{7}}}{\alpha_{_{1}}}+\alpha_{_{1}}\rho$
                                  &$\Lambda=-8\alpha_{_{4}}[2\alpha_{_{7}}$\\
$H^{(\kappa,\tilde{A})}_{_{4}}.X$ &$+[(\alpha_{_{1}}-\alpha_{_{4}})u$                                    & $-2\alpha_{_{4}}\rho +\alpha_{_{5}} y]~dx$
                                  &~~~~$+(\alpha_{_{1}}-\alpha_{_{4}})\rho$\\
                                  &$+\alpha_{_{5}}~e^{-x}+\alpha_{_{6}}]~\partial_{_{u}}$                & $+[e^{x}(-\alpha_{_{1}}u+\alpha_{_{4}}u-\alpha_{_{6}})+\alpha_{5}]dy$
                                  &~~~~$-\frac{2\alpha_{_{4}}\alpha_{_{7}}}{\alpha_{_{1}}}],$\\
                                  &$-(\alpha_{_{1}}~\rho+\alpha_{_{5}}~y+\alpha_{_{7}})~\partial_{_{v}}$ & $+y~e^x~\alpha_{_{4}}~du+(-\alpha_{_{1}}+2\alpha_{_{4}})dv$          & $\tilde{A}=-1+\kappa$\\
\hline
\hline
\end{tabular}}}
\end{center}
$\bullet$~ The first solution is given by
\begin{eqnarray}\nonumber
I&=&-\alpha_{_{1}}\partial_{_{x}}+\alpha_{_{4}}y~\partial_{_{y}}+[(\alpha_{_{1}}-\alpha_{_{4}})u+e^{-x}\alpha_{_{5}}+\alpha_{_{6}}]\partial_{_{u}}\\\nonumber
~~~~&&+[\alpha_{_{1}}(\rho-2\eta^2)-\alpha_{_{5}} y-\alpha_{_{7}}]\partial_{_{v}},\\\nonumber
Z&=&\Big[\alpha_{_{7}}-\frac{2\alpha_{_{4}}\alpha_{_{7}}}{\alpha_{_{1}}}+(\alpha_{_{1}}-2\alpha_{_{4}})(\rho-2\eta^2)+\alpha_{_{5}}y\Big]dx\\\label{2.13}
~~~~&&+\Big[e^x((\alpha_{_{1}}-\alpha_{_{4}})u-\alpha_{_{6}})+\alpha_{_{5}}\Big]dy+\alpha_{_{4}}~e^x~y~du+(-\alpha_{_{1}}+2\alpha_{_{4}})dv.
\end{eqnarray}
The above solution satisfies equations \eqref{2.1}-\eqref{2.7} provided that
\begin{eqnarray}\nonumber
\Lambda &=&-8\alpha_{_{4}}\Big[2\alpha_{_{7}}+(\alpha_{_{1}}-\alpha_{_{4}})(\rho-2\eta^2)-\frac{2\alpha_{_{4}}\alpha_{_{7}}}{\alpha_{_{1}}}\Big],\\\nonumber
\kappa &=&1.
\end{eqnarray}
$\bullet$~Other solution with zero cosmological constant is
\begin{eqnarray}\nonumber
I&=&-\alpha_{_{7}}~\partial_{_{v}},\\\label{2.14}
Z&=&[\frac{x}{4}(1-\kappa^2)+c_{_{2}}]dx,
\end{eqnarray}
for some constant $c_{_{2}}$.
In this manner, one can investigate that other nine deformed WZW models can be solutions of the GSEs.
The deformed backgrounds and related Killing vectors are summarized in Table 1; moreover,
we have listed the vector field $I$ and one-form $Z$ corresponding to each background in Table 2.
\vspace{3mm}
\section{\label{Sec.III} Non-Abelian T-duals of the YB deformed $H_{_{4}}$ WZW backgrounds }

Here we shall obtain the non-Abelian T-dual spaces for the YB deformed WZW models on $H_{_{4}}$ Lie group (Table 1).
Before proceeding, let us give a brief review of the construction of
Poisson-Lie T-dual $\sigma$-models in the presence of spectator fields \cite{Klim1,Klim2}.

Consider now a non-linear $\sigma$-model with $d$ field variables
$x^{M} = (x^{\mu} , y^\alpha)$, where $x^\mu$'s,  $\mu = 1, . . . , dim~G$ are the coordinates of Lie group $G$ acting freely on
the manifold $\M \approx O \times G$, and $y^\alpha,~\alpha = 1, \cdots , d-dim~G$ are
the coordinates of the orbit $O$ of $G$ in  ${\M}$.
Note that the coordinates $y^\alpha$ do not participate in the Poisson-Lie  T-duality transformations
and are therefore called spectator fields \cite{Sfetsos1}.
The corresponding $\sigma$-model action has the form
\vspace{-1mm}
\begin{eqnarray}\label{3.3}
S = \frac{1}{2} \int d\sigma^{+}  d\sigma^{-} \hspace{-6mm}&&\Big[E_{_{ab}}(g,y^{\alpha})~
R_{+}^a \;R_{-}^b + \phi^{{(1)}}_{a \beta}(g,y^{\alpha})  R_{+}^a \partial_{-} y^{\beta}+
\phi^{{(2)}}_{\alpha b}(g,y^{\alpha}) \partial_{+} y^{\alpha} R_{-}^b\nonumber\\
~~&&+\phi_{_{\alpha\beta}}(g,y^{\alpha})
\partial_{+} y^{\alpha} \partial_{-} y^{\beta} \Big].
\end{eqnarray}
where $R_{\pm}^a$ are the components of the right-invariant one-forms which are constructed by means of
an element $g$ of the Lie group $G$ as
\vspace{-1mm}
\begin{eqnarray}\label{3.4}
R_{\pm}  = (\partial_{\pm} g g^{-1})^a ~ T_a = R_{\pm}^a~ T_a=  \partial_{\pm} x^{\mu}~ R_{\mu}^{~a} ~ T_a.
\end{eqnarray}
As shown, the couplings ${{E}_{_{ab}}},\phi^{{(1)}}_{a \alpha}, \phi^{{(2)}}_{\alpha b} $ and $\phi_{_{\alpha\beta}}$ may depend on all variables $x^\mu$ and $y^\alpha$.
The relation of the couplings ${{E}_{_{ab}}}, \phi^{{(1)}}_{a \beta}, \phi^{{(2)}}_{\alpha b} $
and $\phi_{_{\alpha \beta}}$ in \eqref{3.3} has been given as following \cite{Klim1,Klim2,Sfetsos1}
\vspace{-2mm}
\begin{eqnarray}\label{3.6}
{{E}} &=& \big(E^{{-1}}_{0}+ \Pi\big)^{-1},~~~~~~~~~~~
\phi^{{(1)}} = {{E}}~E^{{-1}}_{0}~F^{^{(1)}},\nonumber\\
\phi^{{(2)}}&=& F^{^{(2)}}~ E^{{-1}}_{0}~{{E}},~~~~~~~~~~~~~~~~
\phi = F -F^{^{(2)}}~\Pi~{{E}}~E^{{-1}}_{0}~F^{^{(1)}},
\end{eqnarray}
such that the couplings $E_{0}, F^{^{(1)}}, F^{^{(2)}}$ and $F$ may be at most functions of
the variables $y^{\alpha}$ only.
Furthermore, $\Pi(g)$ defined by $\Pi^{^{ab}}(g) = b^{^{ac}}(g)~ (a^{-1})_{_{c}}^{^{~b}}(g)$ is the Poisson structure on $G$ so that matrices
$a(g)$ and $b(g)$ are defined as follows:
\vspace{-1mm}
\begin{eqnarray}\label{3.7}
g^{-1} T_{{_a}}~ g &=& a_{_{a}}^{^{~b}}(g) ~ T_{{_b}},\nonumber\\
g^{-1} {\tilde T}^{{^a}} g &=& b^{^{ab}}(g)~ T_{{_b}}+(a^{-1})_{_{b}}^{^{~a}}(g)~{\tilde T}^{{^b}}.
\end{eqnarray}

One may define the dual $\sigma$-model for the $d$ field variables ${\tilde x}^{M} =({\tilde x}^{\mu} , y^\alpha)$ similar to \eqref{3.3} by replacing the untilded symbols by tilded ones, where ${\tilde x}^{\mu}$'s parameterize an element ${\tilde g}\in {\tilde G}$, whose dimension is
equal to that of $G$, and the rest of the variables are the same $y^\alpha$'s used in \eqref{3.3}.
The relationship between the couplings of the dual action, ${{{\tilde E}}^{{ab}}}, {\tilde \phi}^{\hspace{0mm}{(1)^{ a}}}_{~~~\beta},
{\tilde \phi}^{\hspace{0mm}{(2)^{ b}}}_{\alpha}$ and ${\tilde \phi}_{_{\alpha \beta}}$  and the original one is given by \cite{Klim1,Klim2,Sfetsos1}
\begin{eqnarray}\label{3.8}
{{\tilde E}} &=& \big(E_{0}+ {\tilde \Pi}\big)^{-1},~~~~~~~~~~
{\tilde \phi}^{{(1)}} =  {{\tilde E}}~F^{^{(1)}},\nonumber\\
{\tilde \phi}^{{(2)}} &=& - F^{^{(2)}} ~{{\tilde E}},~~~~~~~~~~~~~~~~~
{\tilde \phi}           = F-F^{^{(2)}} ~{{\tilde E}}~F^{^{(1)}}.
\end{eqnarray}
Analogously, one can define matrices ${\tilde a} (\tilde g), {\tilde b} (\tilde g)$ and ${\tilde \Pi} (\tilde g)$ by just replacing the untilded symbols
by tilded ones.
In the following, we will apply the procedure mentioned above to construct the non-Abelian T-dual spaces of the YB deformed backgrounds of Table 1.
As an example, we discuss in details the non-Abelian T-dualization of the YB deformed background $H^{(\kappa,\eta)}_{_{4}}.{II}$.
\vspace{3mm}
\subsection{The non-Abelian T-dualization of the background $H^{(\kappa,\eta)}_{_{4}}.II$}\label{Sec.III.1}
\subsubsection{The original model as the deformed one}\label{Sec.III.1.1}
The original model is constructed on $2+2$-dimensional manifold ${\M} \approx O \times G$ in which $G$ is considered to be
the Lie group $A_{_{2}}$ with Lie algebra ${\cal A}_2$, while $O$ is the orbit of $G$ in ${\M}$.
We use the coordinates $\{x_{_{1}}, x_{_{2}}\}$ for the $A_{_{2}}$, and employ $y^\alpha =\{y_{_{1}}, y_{_{2}}\}$ for the orbit $O$.
We shall show the background of original model is equivalent to the YB deformed background $H^{(\kappa,\eta)}_{_{4}}.II$.
The Lie algebra of the semi-Abelian double $({\cal A}_2 , 2{\cal A}_1)$ is defined by the following non-zero Lie brackets
\vspace{-1mm}
\begin{eqnarray}\label{3.9}
[T_{_{1}} , T_{_{2}}]~=~T_{_{2}},~~~~~[T_{_{1}} ~, {\tilde T}^{^{2}}]=-{\tilde T}^{^{2}},~~~~~[T_{_{2}} ~, {\tilde T}^{^{2}}]={\tilde T}^{^{1}}.
\end{eqnarray}
where $\{T_{_{1}} , T_{_{2}}\}$ and $\{{\tilde T}^{^{1}} , {\tilde T}^{^{2}}\}$ are the basis of ${\cal A}_{_{2}}$ and $2{\cal A}_{_{1}}$, respectively.
By parametrization of the group element $A_{_{2}}$ as
\begin{eqnarray}\label{3.10}
g~=e^{x_{_{1}} T_{_{1}}}~e^{x_{_{2}} T_{_{2}}},
\end{eqnarray}
the right invariant one-forms
$R_{\pm}^a$ are derived as follows
\begin{eqnarray}\label{3.11}
R^{1}_{\pm}=\partial_{\pm} x_1,~~~~~~~~~R^{2}_{\pm}=e^{x_1} \partial_{\pm} x_2.
\end{eqnarray}
To achieve a $\sigma$-model with the background $H^{(\kappa,\eta)}_{_{4}}.II$
we choose the spectator-dependent matrices in the following form
\vspace{-1mm}
\begin{eqnarray}
E_{0_{ab}}&=&\left( \begin{tabular}{cc}
$\rho-2\eta^2$ & $-\kappa y_{_{1}}$  \\
$\kappa y_{_{1}}$ & 0 \\	
\end{tabular} \right),\hspace{1cm}F^{(1)}_{a \beta }=\left( \begin{tabular}{cc}
0 & -1 \\
1 & 0 \\
\end{tabular} \right),\nonumber\\
F^{(2)}_{ \alpha b }&=&\left( \begin{tabular}{cc}
0 & 1   \\
-1 & 0 \\	
\end{tabular} \right),\hspace{2.7cm}F_{\alpha \beta }=\left( \begin{tabular}{cc}
0 & 0  \\
0 & 0  \\
\end{tabular} \right).\label{3.12}
\end{eqnarray}
Since the dual Lie group, $2A_{_{1}}$, is assumed to be Abelian, it follows from \eqref{3.7} that $\Pi^{ab}(g)=0$.
Using these and \eqref{3.6} one can construct the action \eqref{3.3} on the manifold ${\M} \approx  O \times G$.
The background including the metric and $B$-field is given as follows:
\vspace{-1.2mm}
\begin{eqnarray}\label{3.12.1}
ds^2 &=& (\rho-2\eta^2)dx_{_{1}}^2-2 dx_{_{1}} dy_{_{2}}+2e^{x_{_{1}}}dx_{_{2}}dy_{_{1}},\\\label{3.12.2}
B &=& \kappa y_{_{1}}~dx_{_{2}}\wedge dx_{_{1}}.
\end{eqnarray}
One can use the transformation $(x,u,y,v)$ instead of $(x_{_{1}},x_{_{2}},y_{_{1}},y_{_{2}})$ to conclude that the above background
is nothing but the YB deformed background $H^{(\kappa,\eta)}_{_{4}}.II$ as was represented in Table 1.
Thus, we showed that the background $H^{(\kappa,\eta)}_{_{4}}.II$
can be considered as the original model of a dual pair of $\sigma$-models related by Poisson-Lie symmetry.
In this manner, one can obtain the spectator-dependent matrices for all backgrounds of Table 1.
The results are summarized in Table 3.
Note that for all models, the matrix $F_{_{\alpha \beta}}$ vanishes.
\subsubsection{The dual model}\label{Sec.III.1.2}
The dual model is constructed on a $2+2$-dimensional manifold
$\tilde {\M} \approx O \times \tilde {G}$ with two-dimensional Abelian Lie group ${\tilde {G}}=2A_{_{1}}$ acting freely on it.
In the same way we parameterize the corresponding Lie group $\tilde {G}$ (Abelian Lie group $2A_{_{1}}$)
with coordinates ${\tilde x}^{^{\mu}} = \{{\tilde x_{_{1}}} , {\tilde x_{_{2}}}\}$
and element of the group as
\vspace{-1mm}
\begin{eqnarray}\label{3.13}
\tilde g=e^{\tilde x_{_{1}} {\tilde T}^{1}}e^{\tilde x_{_{2}} {\tilde T}^{2}},
\end{eqnarray}
then, we have
\vspace{-1mm}
\begin{eqnarray}\label{3.14}
\tilde R_{\pm 1}=\partial_{\pm}\tilde x_1,~~~~~~~~~ \tilde R_{\pm 2}=\partial_{\pm} \tilde  x_2.
\end{eqnarray}
Utilizing relation \eqref{3.7} for untilded quantities, we get
{\small \begin{eqnarray}\label{3.15}
\tilde \Pi_{_{ab}}=\left( \begin{tabular}{cc}
0                   & $-\tilde  x_{_{2}} $  \\
$\tilde  x_{_{2}} $ &    0                   \\
\end{tabular} \right).
\end{eqnarray}}
Now  inserting \eqref{3.12} and \eqref{3.15} into equations \eqref{3.8}
one can obtain dual couplings, giving us
{\small \begin{eqnarray*}
\tilde	E^{ab}=\frac{1}{{\Delta}}\left( \begin{tabular}{cc}
0  & 1  \\
-1  & $\frac{(\rho-2\eta^2)}{\Delta}$  \\	
\end{tabular} \right),~~~~~~~	{\tilde \phi}^{\hspace{0mm}{(1)^{ a}}}_{~~~\beta}=\frac{1}{{\Delta}}\left( \begin{tabular}{cc}
1 & 0  \\
$\frac{(\rho-2\eta^2)}{\Delta}$ & 1 \\
\end{tabular} \right),
\end{eqnarray*}}
{\small \begin{eqnarray}
{\tilde \phi}^{\hspace{0mm}{(2)^{ b}}}_{\alpha}=\frac{1}{{\Delta}}\left( \begin{tabular}{cc}
1 & $\frac{-(\rho-2\eta^2)}{\Delta}$  \\
0 & 1 \\
\end{tabular} \right),~~~~~~~ {\tilde \phi}_{_{\alpha \beta }}=\frac{1}{{\Delta}}\left( \begin{tabular}{cc}
$\frac{-(\rho-2\eta^2)}{\Delta}$ & -1 \\
1 & 0\\
\end{tabular} \right),
\end{eqnarray}}
where ${\Delta}=\kappa y_{_{1}}+\tilde{x}_{_{2}}$.
\\
\\
\begin{center}
\small {{{\bf Table 3.}~The spectator-dependent background matrices of the original models}}
{\scriptsize
\renewcommand{\arraystretch}{1.3}{
\begin{tabular}{lccc} \hline \hline
Symbol &  $E_{_{0}}$  & $F^{(1)}$  & $F^{(2)}$ \\
\hline
\vspace{3mm}
            {$H^{(\kappa)}_{_{4}}.I$} &
            $\begin{pmatrix}
            \rho            & -\kappa y_{_{1}}\\
            \kappa y_{_{1}} & 0
            \end{pmatrix} $
            &
            $\begin{pmatrix}
            0 & -1\\
            1 & 0
            \end{pmatrix} $
            &
            $\begin{pmatrix}
            0  & 1\\
            -1 & 0
            \end{pmatrix} $
             \\
\vspace{3mm}
            {$H^{(\kappa,\eta)}_{_{4}}.II$} &
            $\begin{pmatrix}
            \rho-2\eta^2    & -\kappa y_{_{1}}\\
            \kappa y_{_{1}} & 0
            \end{pmatrix} $
            &
            $\begin{pmatrix}
            0 & -1\\
            1 & 0
            \end{pmatrix} $
            &
            $\begin{pmatrix}
            0  & 1\\
            -1 & 0
            \end{pmatrix} $
            \\
\vspace{3mm}
            {$H^{(\kappa,\eta,\tilde{A})}_{_{4}}.III$} &
            $\begin{pmatrix}
            \rho & 0\\
            0    & -\rho\eta^2
            \end{pmatrix} $
            &
            $\begin{pmatrix}
            0        & -1\\
            1-\kappa & -\tilde{A}
            \end{pmatrix} $
            &
            $\begin{pmatrix}
            0  & 1+\kappa\\
            -1 & \tilde{A}
            \end{pmatrix} $
            \\
\vspace{3mm}
            {$H^{(\kappa,\eta,\tilde{A})}_{_{4}}.IV$}&
            $\begin{pmatrix}
            \frac{\rho}{1-\eta^2}                                       & -\kappa y_{_{1}}+\frac{y_{_{1}}(\tilde{A}-\eta^2)}{1-\eta^2}\\
            \kappa y_{_{1}}-\frac{y_{_{1}}(\tilde{A}+\eta^2)}{1-\eta^2} & 0
            \end{pmatrix} $
            &
            $\begin{pmatrix}
            0 & -\frac{1}{1-\eta^2}\\
            1 & 0
            \end{pmatrix} $
            &
            $\begin{pmatrix}
            0                    & 1\\
            -\frac{1}{1-\eta^2} & 0
            \end{pmatrix} $
            \\
\vspace{3mm}
            {$H^{(\kappa,\eta,\tilde{A})}_{_{4}}.V$}&
            $\begin{pmatrix}
            \rho                                              & -(\kappa+\tilde{A}) y_{_{1}}-\frac{2\eta^2}{1-\eta^2}\\
            (\kappa+\tilde{A}) y_{_{1}}-\frac{2\eta^2}{1-\eta^2} & 0
            \end{pmatrix} $
            &
            $\begin{pmatrix}
            0 & -1\\
            1 & 0
            \end{pmatrix} $
            &
            $\begin{pmatrix}
            0 & 1\\
            -1 & 0
            \end{pmatrix} $
            \\
\vspace{3mm}
            {$H^{(\kappa,\eta,\tilde{A})}_{_{4}}.VI$}&
            $\begin{pmatrix}
            \rho                         & -\frac{\rho\eta^2}{1-\eta^2}\\
            -\frac{\rho\eta^2}{1-\eta^2} & 0
            \end{pmatrix} $
            &
            $\begin{pmatrix}
            0                    & -1\\
            1-(\kappa+\tilde{A}) & -\tilde{A}
            \end{pmatrix} $
            &
            $\begin{pmatrix}
            0  & 1+(\kappa+\tilde{A})\\
            -1 & \tilde{A}
            \end{pmatrix} $
            \\
\vspace{3mm}
            {$H^{(\kappa,\eta,\tilde{A})}_{_{4}}.VII$}&
            $\begin{pmatrix}
            \frac{\rho}{1+\eta^2} & 0\\
            0                     & -\frac{\rho\eta^2}{1+\eta^2}
            \end{pmatrix} $
            &
            $\begin{pmatrix}
            0        & -1\\
            1-\kappa & -\tilde{A}
            \end{pmatrix} $
            &
            $\begin{pmatrix}
            0  & 1+\kappa\\
            -1 & \tilde{A}
            \end{pmatrix} $
            \\
\vspace{3mm}
            {$H^{(\kappa,\eta)}_{_{4}}.VIII$}&
            $\begin{pmatrix}
            \rho(1-\eta^2)          & (-\kappa+\eta^2)y_{_{1}}\\
            (\kappa+\eta^2)y_{_{1}} & 0
            \end{pmatrix} $
            &
            $\begin{pmatrix}
            0 & -(1-\eta^2)\\
            1 & 0
            \end{pmatrix} $
            &
            $\begin{pmatrix}
            0           & 1\\
            -(1-\eta^2) & 0
            \end{pmatrix} $
            \\
\vspace{3mm}
            {$H^{(\kappa,\eta,\tilde{A})}_{_{4,q}}.IX$}&
            ${\begin{pmatrix}
            \frac{\rho(1-\eta^2)}{1-\eta^2q^4} & -\Omega\\
            \Omega                             & 0
            \end{pmatrix} }^\ast$
            &
            $\begin{pmatrix}
            0 & -\frac{(1-\eta^2)}{1-\eta^2q^4}\\
            1 & 0
            \end{pmatrix} $
            &
            $\begin{pmatrix}
            0                               & 1\\
            -\frac{(1-\eta^2)}{1-\eta^2q^4} & 0
            \end{pmatrix} $
            \\
\vspace{3mm}
            {$H^{(\kappa,\tilde{A})}_{_{4}}.X$}&
            $\begin{pmatrix}
            \rho                   & -(\kappa-\tilde{A})y_{_{1}}\\
            (k-\tilde{A})y_{_{1}}  & 0
            \end{pmatrix} $
            &
            $\begin{pmatrix}
            0 & -1\\
            1 & 0
            \end{pmatrix} $
            &
            $\begin{pmatrix}
            0  & 1\\
            -1 & 0
            \end{pmatrix} $
            \\
\hline
\hline
\end{tabular}}}
\end{center}
{\vspace{-2mm}
~~~~~\small$^\ast \Omega=[\kappa-\frac{\tilde{A}q^2(1-\eta^2)+\eta^2(1-q^4)}{1-\eta^2q^4}]y_{_{1}}.$
}\\
\\
Finally, using these one can write down the action of the dual model. The corresponding metric and $\tilde{B}$-field are given by
\begin{eqnarray}
d\tilde{s}^2&=&\frac{1}{{\Delta}} \Big[2 d\tilde{x}{_{_{1}}}~dy_{_{1}}+\frac{\rho-2\eta^2}{{\Delta}}d\tilde{x}^2\hspace{-1.5mm}{_{_{2}}}
+2d\tilde{x}{_{_{2}}}~dy_{_{2}}-\frac{\rho-2\eta^2}{{\Delta}}dy_{_{1}}^2\Big], \label{3.17}\\
\tilde{B}&=&\frac{1}{{\Delta}} \Big[d\tilde{x}{_{_{1}}}\wedge d\tilde{x}{_{_{2}}}+\frac{\rho-2\eta^2}{{\Delta}}d\tilde{x}{_{_{2}}}\wedge dy_{_{1}}+dy_{_{2}} \wedge dy_{_{1}}\Big]. \label{3.18}
\end{eqnarray}

\vspace{-1mm}
\begin{center}
\small {{{\bf Table 4.}~Non-Abelian T-dual backgrounds of the YB deformed $H_{4}$  WZW models}}
{\scriptsize
\renewcommand{\arraystretch}{1.58}{
\begin{tabular}{lll} \hline \hline
Dual symbol &  Background  & Comments \\
\hline
                                    &\hspace{1cm} $d\tilde{s}^2=\frac{1}{\kappa y_{_{1}}+\tilde{x}{_{_2}}}\Big[2d\tilde{x}{_{_{1}}}~dy_{_{1}}+\frac{\rho}{\kappa y_{_{1}}+\tilde{x}{_{_{2}}}}d\tilde{x}^2\hspace{-1.7mm}{_{_{2}}}
                                                                                                                       +2d\tilde{x}{_{_{2}}}dy_{_{2}}-\frac{\rho}{\kappa y_{_{1}}+\tilde{x}{_{_{2}}}}dy_{_{1}}^{2}\Big],$ & \\
$\widetilde{H^{(\kappa)}_{_{4}}.I}$ &\hspace{1.3cm}$\tilde{B}=\frac{1}{\kappa y_{_{1}}+\tilde{x}{_{_{2}}}}\Big[d\tilde{x}{_{_{1}}}\wedge d\tilde{x}{_{_{2}}}+\frac{\rho}{\kappa y_{_{1}}
                                                                                                                       +\tilde{x}{_{_{2}}}}d\tilde{x}{_{_{2}}}\wedge dy_{_{1}}+dy_{_{2}} \wedge dy_{_{1}}\Big]$ &\\
\hline
                                          &\hspace{1cm} $d\tilde{s}^2=\frac{1}{\kappa y_{_{1}}+\tilde{x}{_{_2}}}\Big[2d\tilde{x}{_{_{1}}}~dy_{_{1}}
                                                                    +\frac{\rho-2\eta^2}{\kappa y_{_{1}}+\tilde{x}{_{_{2}}}}d\tilde{x}^2\hspace{-1.7mm}{_{_{2}}}
                                                                    +2d\tilde{x}{_{_{2}}}dy_{_{2}}-\frac{\rho-2\eta^2}{\kappa y_{_{1}}+\tilde{x}{_{_{2}}}}dy_{_{1}}^{2}\Big],$ &\\
$\widetilde{H^{(\kappa,\eta)}_{_{4}}.II}$ &\hspace{1.2cm} $\tilde{B}=\frac{1}{\kappa y_{_{1}}+\tilde{x}{_{_2}}}\Big[d\tilde{x}{_{_{1}}}\wedge d\tilde{x}{_{_{2}}}
                                                                    +\frac{\rho-2\eta^2}{\kappa y_{_{1}}+\tilde{x}{_{_{2}}}}d\tilde{x}{_{_{2}}}\wedge dy_{_{1}}+dy_{_{2}} \wedge dy_{_{1}}\Big]$ &\\
\hline
                            &\hspace{1cm}$d\tilde{s}^2=\frac{1}{\tilde{x}^2\hspace{-1.7mm}{_{_{2}}}-\rho^2\eta^2}\Big\{-\rho \eta^2 d\tilde{x}^2\hspace{-1.7mm}{_{_{1}}}+2\tilde{x}{_{_{2}}}~ d\tilde{x}{_{_{1}}} dy_{_{1}}
                                                                                                                            +\rho~d\tilde{x}^2\hspace{-1.7mm}{_{_{2}}}-2\rho \kappa ~d\tilde{x}{_{_{2}}} dy_{_{1}} $ &\\
                                                     &\hspace{1.4cm} $+2(\tilde{x}{_{_{2}}}-\rho \tilde{A})~d\tilde{x}{_{_{2}}} dy_{_{2}}
                                                                      -\rho(1-\kappa^2)dy_{_{1}}^2+2\kappa(\tilde{A}\rho-\tilde{x}{_{_{2}}})dy_{_{1}}dy_{_{2}}$ &\\
$\widetilde{H^{(\kappa,\eta,\tilde{A})}_{_{4}}.III}$ &\hspace{1.4cm} $+[\rho(\eta^2+\tilde{A}^2)-2\tilde{A}\tilde{x}{_{_{2}}}]dy_{_{2}}^2\Big\},$      &\\
                           &\hspace{1.2cm}$\tilde{B}=\frac{1}{\tilde{x}^2\hspace{-1.7mm}{_{_{2}}}-\rho^2\eta^2}\Big[\kappa\tilde{x}{_{_{2}}}~dy_{_{1}}\wedge d\tilde{x}{_{_{1}}}
                                                                                                 +(\tilde{A}\tilde{x}{_{_{2}}}-\rho\eta^2)dy_{_{2}}\wedge d\tilde{x}{_{_{1}}}$ &\\
                                                    &\hspace{1.4cm} $+(\tilde{A}\rho-\tilde{x}{_{_{2}}})dy_{_{1}}\wedge dy_{_{2}}\Big]$ &\\
\hline
                                                     &\hspace{1cm}$d\tilde{s}^2=\frac{1}{\delta_1}\Big[-2(1-\eta^2) d\tilde{x}{_{_{1}}}dy_{_{1}}+\frac{\rho(1-\eta^2)}{\delta_1}d\tilde{x}^2\hspace{-1.7mm}{_{_{2}}}$
                                                     &\hspace{-0.8cm}$\delta_1=\Big\{[(\kappa-1)\eta^2$\\
$\widetilde{H^{(\kappa,\eta,\tilde{A})}_{_{4}}.IV}$  &\hspace{1.5cm}$-\frac{\rho(1-\eta^2)}{\delta_1}dy_{_{1}}^2
                                                                    -2~d\tilde{x}{_{_{2}}}dy_{_{2}}\Big],$&\hspace{-0.6cm} $+\tilde{A}-\kappa]y_{_{1}}$\\
                                                     &\hspace{1.1cm} $\tilde{B}=\frac{1}{\delta_1}\Big[(1-\eta^2) d\tilde{x}{_{_{2}}}\wedge d\tilde{x}{_{_{1}}}
                                                                              +\frac{\rho(1-\eta^2)}{\delta_1} d\tilde{x}{_{_{2}}}\wedge dy_{_{1}}
                                                                              +dy_{_{1}}\wedge dy_{_{2}}\Big]$ &\hspace{-0.6cm} $-\tilde{x}{_{_{2}}}(1-\eta^2)\Big\}$\\
\hline
                                                  &$d\tilde{s}^2=\frac{1}{\delta_2}\Big[-2(1-\eta^2)d\tilde{x}{_{_{1}}}dy_{_{1}}-\frac{\rho(1-\eta^2)^2}{\delta_2}d\tilde{x}^2\hspace{-1.7mm}{_{_{2}}} $&
                                                  \hspace{-10mm} $\delta_2=\Big\{(\eta^2-1)[(\kappa +\tilde{A})y_{_{1}}$\\
$\widetilde{H^{(\kappa,\eta,\tilde{A})}_{_{4}}.V}$&\hspace{4.1mm} $-2(1-\eta^2)d\tilde{x}{_{_{2}}}dy_{_{2}}+\frac{\rho(1-\eta^2)^2}{\delta_2}dy_{_{1}}^2 \Big],$
                                                  & \hspace{-6mm} $+\tilde{x}{_{_{2}}}]y_{_{1}}-2\eta^2\Big\}$\\
                                                  &\hspace{1.6mm}$\tilde{B}=\frac{1}{\delta_2}\Big[(1-\eta^2) d\tilde{x}{_{_{2}}}\wedge d\tilde{x}{_{_{1}}}
                                                              -\frac{\rho(1-\eta^2)^2}{\delta_2} d\tilde{x}{_{_{2}}}\wedge dy_{_{1}}+(1-\eta^2)dy_{_{1}}\wedge dy_{_{2}}\Big]$ & \\
\hline                                   & $d\tilde{s}^2=\frac{1}{\delta_{_{3}}^{+}\delta_{_{3}}^{-}}\Big\{[-2\rho \eta^2(\eta^2-1)]d\tilde{x}{_{_{1}}}d\tilde{x}{_{_{2}}}
                                                  +[\rho(\eta^2-1)]d\tilde{x}^2\hspace{-1.7mm}{_{_{2}}} $ &  \\
                                   &\hspace{4.5mm} $+\Big[2(\eta^2-1)(\tilde{x}{_{_{2}}}(\eta^2-1)+(\kappa+\tilde{A})\rho\eta^2) \Big]d\tilde{x}{_{_{1}}}dy_{_{1}}$ & $~~~~~~$\\
                                   &\hspace{4.5mm} $+\Big[2\tilde{A}(\eta^2-1)\rho\eta^2\Big]d\tilde{x}{_{_{1}}}dy_{_{2}}-\Big[2\rho(\eta^2-1)^2(\kappa+\tilde{A}) \Big]d\tilde{x}{_{_{2}}}dy_{_{1}}$ &\\
                                   &\hspace{4.5mm} $-\Big[2(\eta^2-1)((-\tilde{x}{_{_{2}}}+\tilde{A}\rho)\eta^2-\rho \tilde{A}+\tilde{x}{_{_{2}}}) \Big]d\tilde{x}{_{_{2}}}dy_{_{2}}$ &\\
$\widetilde{H^{(\kappa,\eta,\tilde{A})}_{_{4}}.VI}$ &\hspace{4.5mm} $-\Big[\rho(\eta^2-1)^2((\kappa+\tilde{A})^2-1) \Big]dy_{_{1}}^2$ &
                                                      \hspace{-1cm} $\delta_{_{3}}^{\pm}=\tilde{x}{_{_{2}}}(\eta^2-1)\pm \rho \eta^2 $\\
                                   &\hspace{4.5mm} $-\Big\{2(\eta^2-1)\Big[-\rho \eta^2+((-\tilde{x}{_{_{2}}}+\tilde{A}\rho)\eta^2-\rho\tilde{A}+\tilde{x}{_{_{2}}})(\kappa+\tilde{A})\Big] \Big\} dy_{_{1}} dy_{_{2}}$ &\\
                                   &\hspace{4.5mm} $-\Big[(\eta-1)^2(\eta+1)^2 \tilde{A} (\rho\tilde{A}-2\tilde{x}{_{_{2}}})\Big]dy_{_{2}}^2 \Big\},$ &\\

                        &\hspace{1.5mm} $\tilde{B}=\frac{1}{\delta_{_{3}}^{+}\delta_{_{3}}^{-}}\Big\{\tilde{x}{_{_{2}}}(\eta^2-1)^2~d\tilde{x}{_{_{1}}}\wedge  \tilde{x}{_{_{2}}}$ &\\
                                   &\hspace{4.5mm} $+\Big[(\eta^2-1)(\rho \eta^2+(k+\tilde{A})(\tilde{x}{_{_{2}}}(\eta^2-1)) \Big]dy_{_{1}} \wedge d\tilde{x}{_{_{1}}}$ & \\
                                   &\hspace{4.5mm} $+\Big[\tilde{A}\tilde{x}{_{_{2}}}(\eta^2-1)^2 \Big] dy_{_{2}} \wedge d\tilde{x}{_{_{1}}}
                                          +\rho (\eta^2-1)^2~d\tilde{x}{_{_{2}}} \wedge dy_{_{1}}+\rho\eta^2 (\eta^2-1)~d\tilde{x}{_{_{2}}} \wedge dy_{_{2}}$ &\\
                                   &\hspace{4.5mm} $+\Big[(\eta^2-1)[-\rho \eta^2(\kappa+\tilde{A})+((-\tilde{x}{_{_{2}}}+\tilde{A}\rho)\eta^2-\rho\tilde{A}+\tilde{x}{_{_{2}})}] \Big] \Big\}dy_{_{1}} \wedge dy_{_{2}}$ &\\
\hline
\hline
\end{tabular}}}
\end{center}
\begin{center}
\small {{{\bf Table 4.}~Continued.}}
{\scriptsize
\renewcommand{\arraystretch}{1.58}{
\begin{tabular}{lll} \hline \hline
Dual symbol &  Background  & Comments \\
\hline
                                    & $d\tilde{s}^2=\frac{1}{\delta_4}\Big\{-\rho\eta^2(\eta^2+1)d\tilde{x}^2\hspace{-1.7mm}{_{_{1}}}+2\tilde{x}{_{_{2}}}(\eta^2+1)^2 d\tilde{x}{_{_{1}}}dy_{_{1}}
                                                   +\rho(\eta^2+1)d\tilde{x}^2\hspace{-1.7mm}{_{_{2}}} $ & \\
                                    &\hspace{4.5mm} $-2\rho \kappa (\eta^2+1)d\tilde{x}{_{_{2}}}dy_{_{1}}-2[(\eta^2+1)((-\eta^2-1)\tilde{x}{_{_{2}}}
                                     +\rho \tilde{A})] d\tilde{x}{_{_{2}}}dy_{_{2}}$ & \\
                                    &\hspace{4.5mm} $+\rho(\eta^2+1)(\kappa^2-1)dy_{_{1}}^2+2\kappa(\eta^2+1)\Big[(-\eta^2-1)\tilde{x}{_{_{2}}}
                                      +\rho\tilde{A} \Big]dy_{_{1}} dy_{_{2}}$ & \hspace{-8mm}$\delta_4=\Big\{-\rho^2\eta^2$\\
$\widetilde{H^{(\kappa,\eta,\tilde{A})}_{_{4}}.VII}$ &\hspace{4.5mm} $+\Big[(\eta^2+1)(-2\tilde{A}\tilde{x}{_{_{2}}}\eta^2+\tilde{A}^2\rho+\eta^2\rho-2\tilde{x}{_{_{2}}}\tilde{A}) \Big] dy_{_{2}}^2 \Big\},$ &
                                                        \hspace{-5mm}$+\tilde{x}{_{_{2}}}^2(\eta^2+1)^2\Big\}$\\

                        &\hspace{1.5mm} $\tilde{B}=\frac{1}{\delta_4}\Big[\tilde{x}{_{_{2}}}(\eta^2+1)^2 d\tilde{x}{_{_{1}}}\wedge d\tilde{x}{_{_{2}}}
                                    +\kappa \tilde{x}{_{_{2}}}(\eta^2+1)^2 dy_{_{1}} \wedge d\tilde{x}{_{_{1}}}\Big]$ & \\
                                   &\hspace{4.5mm} $+\Big[(\eta^2+1)(\tilde{A}\eta^2\tilde{x}{_{_{2}}}-\eta^2 \rho+\tilde{A}\tilde{x}{_{_{2}}}) \Big] dy_{_{2}} \wedge d\tilde{x}{_{_{1}}}$ & \\
                                   &\hspace{4.5mm} $+\rho(\eta^2+1)d\tilde{x}{_{_{2}}}\wedge dy_{_{1}}+(\eta^2+1)((-\eta^2-1)\tilde{x}{_{_{2}}}+\rho \tilde{A})dy_{_{1}} \wedge dy_{_{2}} \Big] $ &\\
\hline
                                     & $d\tilde{s}^2=\frac{1}{\delta_{_{5}}^{+}\delta_{_{5}}^{-}}\Big[-2\eta^2 y_{_{1}} d\tilde{x}{_{_{1}}}d\tilde{x}{_{_{2}}}
                                                    +2(\kappa y_{_{1}}+\tilde{x}{_{_{2}}}) d\tilde{x}{_{_{1}}}dy_{_{1}}+\rho (1-\eta^2) d\tilde{x}^2\hspace{-1.7mm}{_{_{2}}} $
                                     &\hspace{-8mm} $\delta_{_{5}}^{\pm}=(\kappa \pm \eta^2)y_1+\tilde{x}{_{_{2}}}$\\
$\widetilde{H^{(\kappa,\eta)}_{_{4}}.VIII}$& \hspace{4.5mm}$+\rho (\eta^2-1) dy_{_{1}}^2+2[(1-\eta^2)(\kappa y_{_{1}}+\tilde{x}{_{_{2}}})]d\tilde{x}{_{_{2}}} dy_{_{2}}
                                                  +2\eta^2 y_{_{1}}(\eta^2-1) dy_{_{1}} dy_{_{2}} \Big],$ &\\

                        &\hspace{1.5mm}$\tilde{B}=\frac{1}{\delta_{_{5}}^{+}\delta_{_{5}}^{-}}\Big[(\kappa y_{_{1}}+\tilde{x}{_{_{2}}})d\tilde{x}{_{_{1}}}\wedge d\tilde{x}{_{_{2}}}
                                    +\eta^2 y_{_{1}} dy_{_{1}} \wedge d\tilde{x}{_{_{1}}}+\rho(\eta^2-1)dy_{_{1}} \wedge d\tilde{x}{_{_{2}}} $ & \\
                                    &\hspace{4.5mm} $+[(1-\eta^2)\eta^2y_{_{1}}]d\tilde{x}{_{_{2}}}\wedge dy_{_{2}}+[(\eta^2-1)(\kappa y_{_{1}}+\tilde{x}{_{_{2}}})]dy_{_{1}} \wedge dy_{_{2}} \Big]$ & \\
\hline
                                     & $d\tilde{s}^2=\frac{1}{\delta_{_{6}}^{+}\delta_{_{6}}^{-}}
                                     \Big\{ 2(1-\eta^2 q^4)(-1+q^4)\eta^2 y_{_{1}}\Big[d\tilde{x}{_{_{1}}} d\tilde{x}{_{_{2}}}+d\tilde{x}{_{_{1}}} dy_{_{1}} \Big]$ & \\
                                     &\hspace{4.5mm} $+\rho(\eta^2-1)(\eta^2 q^4-1)\Big[ d\tilde{x}^2\hspace{-1.7mm}{_{_{2}}}+2d\tilde{x}{_{_{2}}}dy_{_{1}}\Big] $ & \\
                                     &\hspace{5.5mm}$+2(1-\eta^2)\{[(-\kappa q^4+\tilde{A}q^2)\eta^2-\tilde{A}q^2+\kappa]y_{_{1}}+\tilde{x}{_{_{2}}}(1-\eta^2q^4)\}d\tilde{x}{_{_{2}}}dy_{_{2}}$
                                     &\hspace{-8mm} $\delta_{_{6}}^{\pm}=\Big\{ [\pm 1+ (-\kappa\mp1)q^4 $\\
$\widetilde{H^{\kappa,\eta,\tilde{A}}_{_{4,q}}.IX}$ &\hspace{4.5mm} $-[2(1-\eta^2)(1-q^4)\eta^2y_{_{1}}]~dy_{_{1}} dy_{_{2}} \Big\},$ &\hspace{-3mm} $+\tilde{A}q^2]\eta^2-\tilde{A}q^2+\kappa\Big\}y_{_{1}}$\\

                        &\hspace{1.5mm} $\tilde{B}=\frac{1}{\delta_{_{6}}^{+}\delta_{_{6}}^{-}}\Big\{(1-\eta^2q^4)\Big[[(-\kappa q^4+\tilde{A})\eta^2 -\tilde{A} q^2 +\kappa]y_{_{1}}$
                                    &\hspace{-2mm}$-\eta^2 q^4 \tilde{x}{_{_{2}}}+\tilde{x}{_{_{2}}}$ \\
                                    &\hspace{4.5mm} $-\eta^2 q^4 \tilde{x}{_{_{2}}}+\tilde{x}{_{_{2}}} \Big] \Big[ d\tilde{x}{_{_{1}}}\wedge d\tilde{x}{_{_{2}}}
                                                                         +d\tilde{x}{_{_{1}}}\wedge dy_{_{1}}+dy_{_{2}}\wedge dy_{_{1}} \Big] $ & \\
                                    &\hspace{4.5mm} $+(1-\eta^2)[(1-q^4)\eta^2]y_{_{1}} d\tilde{x}{_{_{2}}}\wedge dy_{_{2}} \Big\}$ &\\
\hline
$\widetilde{H^{(\kappa,\tilde{A})}_{_{4}}.X}$ & $d\tilde{s}^2=\frac{1}{\delta_7}\Big[-2d\tilde{x}{_{_{1}}}dy_{_{1}}+\frac{\rho}{\delta_7}d\tilde{x}^2\hspace{-1.7mm}{_{_{2}}}-2d\tilde{x}{_{_{2}}}dy_{_{2}}
                                              -\frac{\rho}{\delta_7}dy_{_{1}}^2\Big],$
                                              &\hspace{-8mm} $\delta_7=(\tilde{A}-\kappa)y_{_{1}}-\tilde{x}{_{_{2}}}$\\

                        &\hspace{1.2mm} $\tilde{B}=\frac{1}{\delta_7}\Big[d\tilde{x}{_{_{2}}}\wedge d\tilde{x}{_{_{1}}}+\frac{\rho}{\delta_7}d\tilde{x}{_{_{2}}}\wedge dy_{_{1}}+dy_{_{1}} \wedge dy_{_{2}}\Big] $ & \\
\hline
\hline
\end{tabular}}}
\end{center}
Then, by using the Poisson Lie T-duality approach in the presence of spectator fields we obtain the backgrounds of dual models for all YB deformed models of Table 1;
the results are summarized in Table 4.
\section{\label{Sec.IV} Integrability of the T-dual $\sigma$-models}
In order to investigate of the integrability of the dual backgrounds, let us give a short review on the method of integrability of the $\sigma$-models
presented by Mohammedi in \cite{Mohammedi}.
In section \ref{Sec.III} we showed that the YB deformed backgrounds of the $H_{4}$ WZW model are equivalent to
original model ones of the non-Abelian T-dual $\sigma$-models. As mentioned in ref. \cite{magr.vicedo.nucl.},
the YB deformed WZW models are integrable, so our original models as YB deformed ones are integrable. In what follows, we only examine the
integrability of the dual backgrounds of Table 4.

\subsection{A short review of the integrability of $\sigma$-model}\label{Sec.IV.1}
Consider the following sigma model action\footnote{As in refs. \cite{Mohammedi2} and \cite{Mohammedi}, instead of the use of the world sheet
coordinates $\tau$ and $\sigma$, we will use the complex coordinates ($z=\sigma+i\tau,\bar{z}=\sigma-i\tau$)
with $\partial=\frac{\partial}{\partial z}$ and $\bar{\partial}=\frac{{\partial}}{\partial\bar{z}}$.
We also adopt the convention that the Levi-Civita tensor $\epsilon^{z\bar{z}}=1$; so we will not have $i$ in front of $B$-field in \eqref{4.1}.}
\vspace{-2mm}
\begin{equation}\label{4.1}
S=\int_{\Sigma}\hspace{0.9mm}dzd\bar{z}(G_{_{MN}}(x)+B_{_{MN}}(x))\hspace{0mm}\partial\hspace{0mm}x^{M}\bar{\partial}x^{N},
\end{equation}
where $x^{M}(z,\bar{z})$ $(M=1,2,...,d)$ are coordinates of the manifold $\M$ so that $G_{_{MN}}$ and $B_{_{MN}}$ are the metric and anti-symmetric tensor fields on $\M$;
furthermore, the $(z,\bar{z})$ are coordinates of the world-sheet $\Sigma$.
The equations of motion for this model can be expressed as \cite{Mohammedi}
\begin{equation}\label{4.2}
\bar{\partial} \partial x^{R}+\Omega^{R}_{MN}~\partial x^{M}~\bar{\partial}x^{N}=0,
\end{equation}
where
\vspace{-2mm}
\begin{equation}\label{4.3}
\Omega^{R}_{_{MN}}=\Gamma^{R}\hspace{-1mm}_{_{MN}}-H^{R}_{_{MN}},
\end{equation}
in which $\Gamma^{R}\hspace{-1mm}_{_{MN}}$ are the
Christoffel coefficients and the components of the field strength are given by
\vspace{-2mm}
\begin{equation}\label{4.4}
H^{R}\hspace{0cm}_{_{MN}}=\frac{1}{2}G^{RS}(\partial_{_{S}}B_{_{MN}}+\partial_{_{N}}B_{_{SM}}+\partial_{_{M}}B_{_{NS}}).
\end{equation}
One can construct a linear system whose consistency conditions are equivalent to the equations of motion \eqref{4.2} as follows \cite{Mohammedi}:
\vspace{-2mm}
\begin{eqnarray}\label{4.5}
[\partial+\partial\hspace{0cm}x^{M}\alpha_{_{M}}(x)]\psi &=& 0,\\\label{4.6}
[\bar{\partial}+\bar{\partial}x^{M}\beta_{_{M}}(x)]\psi  &=& 0,
\end{eqnarray}
where the matrices $\alpha_{_{M}}$ and $\beta_{_{M}}$ are functions of the coordinates $x^{M}$.
The compatibility condition of this linear system yields the
equations of motion, provided that the matrices $\alpha_{_{M}}(x)$
and $\beta_{_{M}}(x)$ must satisfy the following relations \cite{Mohammedi}\
\vspace{-1.5mm}
\begin{equation}\label{4.7}
\partial_{_{M}}\beta_{_{N}}-\partial_{_{N}}\alpha_{_{M}}+[\alpha_{_{M}},\beta_{_{N}}]=\Omega^{R}_{_{MN}}\mu_{_{R}},
\end{equation}
with
\vspace{-2mm}
\begin{equation}\label{4.8}
\beta_{_{M}}-\alpha_{_{M}}=\mu_{_{M}},
\end{equation}
such that the equation \eqref{4.7} can then be rewritten as
\vspace{-2mm}
\begin{equation}\label{4.9}
F_{_{MN}}=-(\nabla_{_{M}}\mu_{_{N}}-\Omega^{R}_{_{MN}}\mu_{_{R}}),
\end{equation}
where the field strength $F_{_{MN}}$ and covariant derivative corresponding to the matrices $\alpha_{_{M}}$ are given as follows:
\vspace{-2mm}
\begin{equation}\label{4.10}
F_{_{MN}}=\partial_{_{M}}\alpha_{_{N}}-\partial_{_{N}}\alpha_{_{M}}+[\alpha_{_{M}},\alpha_{_{N}}],\hspace{1cm}
\nabla_{_{M}}X=\partial_{_{M}}X+[\alpha_{_{M}},X].
\end{equation}
It seems to be of interest to write down symmetric and anti-symmetric parts of \eqref{4.9}, giving \cite{Mohammedi}
\vspace{-2mm}
\begin{eqnarray}
0&=&\nabla_{_{M}}\mu_{_{N}}+\nabla_{_{N}}\mu_{_{M}}-2~\Gamma^{R}\hspace{-1mm}_{_{MN}}\mu_{_{R}},\label{4.11}\\
F_{_{MN}}&=&-\frac{1}{2}(\nabla_{_{M}}\mu_{_{N}}-\nabla_{_{N}}\mu_{_{M}})-H^{R}\hspace{-1mm}_{_{MN}}~\mu_{_{R}}.\label{4.12}
\end{eqnarray}
In this manner, the integrability condition of the $\sigma$-model \eqref{4.1} is equivalent to finding the
matrices $\alpha_{_{M}}$ and $\mu_{_{M}}$ such that they satisfy in \eqref{4.9} (or \eqref{4.11} and \eqref{4.12}).
\vspace{-1mm}
\subsection{Integrability of the non-Abelian T-dual models}\label{Sec.IV.2}
The subject of this section is that investigating the integrability of the dual backgrounds given in Table 4.
Since we are dealing with the non-Abelian T-duality, the Lie group of the dual target manifold is considered to be Abelian Lie group $4 A_{_{1}}$.
Accordingly, the right-invariant one-forms on the $4 A_{_{1}}$
are ${\tilde R}_{_{\pm_{a}}} =\partial_{_{\pm}} {\tilde X}^{^M} {\tilde R}_{_{_{Ma}}}= \partial_{_{\pm}} {\tilde x}_{_{a}}$,
in which ${\tilde x}_{a},~ a=1, \cdots, 4,$ stand for the coordinates of the $4 A_{_{1}}$.
In order to investigate the integrability of the dual $\sigma$-model, the corresponding linear system is taken to have the form \eqref{4.5}-\eqref{4.6} by replacing the untilded symbols by tilded ones.
One may consider the following expansions for matrices ${\tilde \alpha}_{_{M}}$ and ${\tilde \mu}_{_{M}}$ \cite{{eghbali.fortsch},{Rez}}
\vspace{-2mm}
\begin{eqnarray}
\tilde{\alpha}_{_{M}} = {\tilde R}_{_{Ma}} ~ {\tilde A^{^a}}_{_{~b}}({\tilde x})~ {\tilde T}^{^b},\hspace{5mm}
\tilde{\mu}_{_{M}} ={\tilde R}_{_{Ma}} ~ {\tilde B^{^a}}_{_{~b}}({\tilde x})~ {\tilde T}^{^b}.\label{4.13}
\end{eqnarray}
Here we have assumed that the ${\tilde A^{^a}}_{_{~b}}({\tilde x})$ and ${\tilde B^{^a}}_{_{~b}}({\tilde x})$ are not constant and depend on the coordinates of the group manifold.
Inserting the above expansion into equations \eqref{4.11} and \eqref{4.12} and using the fact that
$\tilde f^{ab}_{\hspace{1.2mm}c} =0$, one gets
\vspace{-0.5mm}
\begin{eqnarray}\label{4.14}
\tilde R_{_{Ma}} \tilde {\partial}_{_{N}} {\tilde B^{^a}}_{_{~b}}({\tilde x}) + {\tilde R}_{_{Na}}  \partial_{_{M}} {\tilde B^{^a}}_{_{~b}}({\tilde x})
-2 {\tilde \Gamma}^{^{P}}_{~_{MN}} {\tilde R}_{_{Pa}} {\tilde B^{^a}}_{_{~b}}({\tilde x}) &=& 0,\\\label{4.15}
\big[\partial_{_M} {\tilde A^{^a}}_{_{~b}}({\tilde x}) +\frac{1}{2} \partial_{_M} {\tilde B^{^a}}_{_{~b}}({\tilde x})\big]
{\tilde R}_{_{Na}} -{\tilde R}_{_{Ma}} \big[\partial_{_N} {\tilde A^{^a}}_{_{~b}}({\tilde x}) +\frac{1}{2} \partial_{_N} {\tilde B^{^a}}_{_{~b}}({\tilde x})\big]\\\nonumber
+{\tilde H}^{^{P}}_{~_{MN}} {\tilde R}_{_{Pa}} {\tilde B^{^a}}_{_{~b}}({\tilde x}) &=& 0.
\end{eqnarray}
Now one must try to obtain the solutions of the above equations for the dual backgrounds of Table 4.
Below, as an example we discuss in details the integrability of the dual background $\widetilde{H^{(\kappa,\eta)}_{_{4}}.II}$.
\vspace{-3mm}
\subsubsection{An example: examining the integrability of the dual background $\widetilde{H^{(\kappa,\eta)}_{_{4}}.II}$}\label{Sec.IV.2.1}
First we must calculate the Christoffel symbols ${\tilde \Gamma}^{^{P}}_{~_{MN}}$ and strength field ${\tilde H}^{^{P}}_{~_{MN}}$ correspond to the
metric ${\tilde G}_{_{MN}}$ and anti-symmetric tensor ${\tilde B}_{_{MN}}$ of the background $\widetilde{H^{(\kappa,\eta)}_{_{4}}.II}$ given by the equations \eqref{3.17} and \eqref{3.18}.
The results are given as follows:
\vspace{-1mm}
\begin{eqnarray}\nonumber
{{\tilde \Gamma}^{^{\tilde{x}_{_{1}}}}_{~_{{\tilde{x}_{_{1}}}{\tilde{x}_{_{2}}}}}} &=& -\kappa~{{\tilde \Gamma}^{^{\tilde{x}_{_{1}}}}_{~_{{\tilde{x}_{_{2}}}{y_{_{2}}}}}}=
\kappa~{{\tilde \Gamma}^{^{\tilde{x}_{_{2}}}}_{~_{{\tilde{x}_{_{2}}}{y_{_{1}}}}}}   =   {{\tilde \Gamma}^{^{y_{_{1}}}}_{~_{{\tilde{x}_{_{2}}}{y_{_{1}}}}}}               =
-{{\tilde \Gamma}^{^{y_{_{2}}}}_{~_{{\tilde{x}_{_{1}}}{y_{_{1}}}}}}                 =\kappa  {{\tilde \Gamma}^{^{y_{_{2}}}}_{~_{{y_{_{1}}}{y_{_{2}}}}}} = -\frac{1}{2(\tilde{x}_{_{2}}+\kappa y_{_{1}})},\\\nonumber
{{\tilde \Gamma}^{^{\tilde{x}_{_{2}}}}_{~_{{\tilde{x}_{_{2}}}{\tilde{x}_{_{2}}}}}}  &=& \kappa {{\tilde \Gamma}^{^{y_{_{1}}}}_{~_{{y_{_{1}}}{y_{_{1}}}}}}=-\frac{1}{\tilde{x}_{_{2}}+\kappa y_{_{1}}},~~~
{{\tilde \Gamma}^{^{\tilde{x}_{_{1}}}}_{~_{{\tilde{x}_{_{2}}}{\tilde{x}_{_{2}}}}}}  =-\frac{1}{2}{{\tilde \Gamma}^{^{y_{_{2}}}}_{~_{{\tilde{x}_{_{2}}}{y_{_{1}}}}}}=
\frac{\kappa(\rho-2\eta^2)}{(\tilde{x}_{_{2}}+\kappa y_{_{1}})^2} , \\\label{4.16}
{{\tilde \Gamma}^{^{\tilde{x}_{_{1}}}}_{~_{{\tilde{x}_{_{2}}}{y_{_{1}}}}}}          &=&-2 {{\tilde \Gamma}^{^{y_{_{2}}}}_{~_{{y_{_{1}}}{y_{_{1}}}}}}=\frac{(\rho-2\eta^2)}{2(\tilde{x}_{_{2}}+\kappa y_{_{1}})^2},
\end{eqnarray}
and
\vspace{-1mm}
\begin{eqnarray}\nonumber
\kappa{\tilde H^{^{{\tilde{x}_{_{1}}}}}_{~_{{\tilde{x}_{_{1}}} {\tilde{x}_{_{2}}}}}} &=& {\tilde H^{^{{\tilde{x}_{_{1}}}}}_{~_{{\tilde{x}_{_{2}}} {y_{_{2}}}}}}
= -{\tilde H^{^{{\tilde{x}_{_{2}}}}}_{~_{{\tilde{x}_{_{2}}} {y_{_{1}}}}}}=
\kappa{\tilde H^{^{{y_{_{1}}}}}_{~_{{\tilde{x}_{_{2}}} {y_{_{1}}}}}} =-\kappa {\tilde H^{^{{y_{_{2}}}}}_{~_{{\tilde{x}_{_{1}}} {y_{_{1}}}}}}
=-{\tilde H^{^{{y_{_{2}}}}}_{~_{{y_{_{1}}} {y_{_{2}}}}}}=-\frac{1}{\tilde{x}_{_{2}}+\kappa y_{_{1}}},\\\label{4.17}
{\tilde H^{^{{\tilde{x}_{_{1}}}}}_{~_{{\tilde{x}_{_{2}}} {y_{_{1}}}}}} &=& \frac{\kappa(2\eta^2-1)}{(\tilde{x}_{_{2}}+\kappa y_{_{1}})^2},\hspace{8mm}
{\tilde H^{^{{y_{_{2}}}}}_{~_{{\tilde{x}_{_{2}}} {y_{_{1}}}}}} = \frac{2(\eta^2-2)}{(\tilde{x}_{_{2}}+\kappa y_{_{1}})^2}.
\end{eqnarray}

\begin{center}
\small {{{\bf Table 5.}~Integrability of the non-Abelian T-dual models}}
{\scriptsize
\renewcommand{\arraystretch}{1.3}{
\begin{tabular}{lccc} \hline \hline
Dual symbol &  $\tilde{A}_{a}^{~b}(\tilde{X})$  & $\tilde{B}_{a}^{~b}(\tilde{X})$ & Comments \\
\hline		
            $\widetilde{H^{(\kappa)}_{_{4}}.I}$&
            $\begin{pmatrix}
            a_{_{11}}(\tilde{x}_{_{1}}) & a_{_{12}}(\tilde{x}_{_{1}}) & a_{_{13}}(\tilde{x}_{_{1}}) & a_{_{14}}(\tilde{x}_{_{1}})\\
            \xi_{_{1}}                  & \xi_{_{1}}                  & \xi_{_{1}}                  & \xi_{_{1}}                 \\
            a_{_{31}}(\tilde{x}_{_{2}}) & a_{_{32}}(\tilde{x}_{_{2}}) & a_{_{33}}(\tilde{x}_{_{2}}) & a_{_{34}}(\tilde{x}_{_{2}})\\
            a_{_{41}}(y_{_{2}})         & a_{_{42}}(y_{_{2}})         & a_{_{43}}(y_{_{2}})         & a_{_{44}}(y_{_{2}})        \\
            \end{pmatrix} $
            &
            $\begin{pmatrix}
            0              & 0              & 0               & 0             \\
            \gamma_{_{1}}  & \gamma_{_{1}}  & \gamma_{_{1}}   & \gamma_{_{1}} \\
            \lambda_{_{1}} & \lambda_{_{1}} & \lambda_{_{1}}  & \lambda_{_{1}}\\
            0              & 0              & 0               & 0             \\
            \end{pmatrix} $
            &
            $
            \begin{matrix}
            \xi_{_{1}}=\frac{-3c_{_{1}}(1+\kappa)-a_{_{0}}(2+\kappa)}{2(\tilde{x}_{_{2}}+\kappa y_{_{1}})},\\
            \gamma_{_{1}}=\frac{c_{_{1}}}{\tilde{x}_{_{2}}+\kappa y_{_{1}}},\\
            \lambda_{_{1}}=\frac{-c_{_{1}}+a_{_{0}}}{\tilde{x}_{_{2}}+\kappa y_{_{1}}}\\
            \end{matrix}
            $
             \\
\hline
            $\widetilde{H^{(\kappa,\eta)}_{_{4}}.II}$&
            $\tilde{A}_{_{2}}$
            &
            $\begin{pmatrix}
            0              & 0              & 0              & 0             \\
            \gamma_{_{2}}  & \gamma_{_{2}}  & \gamma_{_{2}}  & \gamma_{_{2}} \\
            \lambda_{_{2}} & \lambda_{_{2}} & \lambda_{_{2}} & \lambda_{_{2}}\\
            0              & 0              & 0              & 0             \\
            \end{pmatrix} $
            &
            $
            \begin{matrix}
            \xi_{_{2}}=\frac{-3c_{_{1}}(1+\kappa)-a_{_{0}}(1+2~\kappa)}{2\kappa(\tilde{x}_{_{2}}+\kappa y_{_{1}})},\\
            \gamma_{_{2}}=\frac{c_{_{1}}}{\tilde{x}_{_{2}}+\kappa y_{_{1}}},\\
            \lambda_{_{2}}=\frac{-c_{_{1}}+a_{_{0}}}{\tilde{x}_{_{2}}+\kappa y_{_{1}}}\\
            \end{matrix}
            $
            \\
\hline
            $\widetilde{H^{(\kappa,\eta,\tilde{A})}_{_{4}}.III}$&
            $\begin{pmatrix}
            a_{_{11}}(\tilde{x}_{_{1}}) & a_{_{12}}(\tilde{x}_{_{1}}) & a_{_{13}}(\tilde{x}_{_{1}}) & a_{_{14}}(\tilde{x}_{_{1}})\\
            \xi_{_{3}}                  & \xi_{_{3}}                  & \xi_{_{3}}                  & \xi_{_{3}}                 \\
            a_{_{31}}(y_{_{1}})         & a_{_{32}}(y_{_{1}})         & a_{_{33}}(y_{_{1}})         & a_{_{34}}(y_{_{1}})        \\
            \gamma_{_{3}}               & \gamma_{_{3}}               & \gamma_{_{3}}               & \gamma_{_{3}}              \\
            \end{pmatrix} $
            &
            $\lambda_{_{3}} C$
            &
            $
            \begin{matrix}
            \vspace{2mm}
            \xi_{_{3}}=-\frac{y_{_{1}} c_{_{1}}}{\tilde{x}_{_{2}}}(\frac{1}{2}-\kappa)+c_{_{3}}(\tilde{x}_{_{2}}),\\
            \gamma_{_{3}}=-\frac{\tilde{A} c_{_{1}}}{\tilde{x}_{_{2}}}+c_{_{2}}(y_{_{2}}),\\
            \lambda_{_{3}}=\frac{c_{_{1}}}{\tilde{x}_{_{2}}}\\
            \end{matrix}
            $
             \\
\hline
            $\widetilde{H^{(\kappa,\eta,\tilde{A})}_{_{4}}.IV}$&
            $\tilde{A}_{_{4}} $
            &
            $\gamma_{_{4}} C $
            &
            $
            \begin{matrix}
            \vspace{2mm}
            \xi_{_{4}}=-\frac{c_{_{1}}(1-2\tilde{A}+\eta^2-2\kappa(\eta^2-1))}{2\sigma(\tilde{x}_{_{2}}(\eta^2-1)+\sigma y_{_{1}})},\\
            \gamma_{_{4}}=c_{_{1}}[\tilde{x}_{_{2}}(\eta^2-1)+\sigma y_{_{1}}]^{-1},\\
            \sigma=\tilde{A}+\kappa(\eta^2-1)-\eta^2\\
            \end{matrix}
            $
            \\
\hline
            $\widetilde{H^{(\kappa,\eta,\tilde{A})}_{_{4}}.V}$&
            $\tilde{A}_{_{5}}$
            &
            $\gamma_{_{5}} C, $
            &
            $
            \begin{matrix}
            \vspace{2mm}
            \xi_{_{5}}=\frac{c_{_{1}}[1+2(\tilde{A}+\kappa)]}{2\sigma(\tilde{A}+\kappa)}+c_{_{2}}(\tilde{x}_{_{2}}),\\
            \gamma_{_{5}}=c_{_{1}}/\sigma,\\
            \sigma=(\eta^2-1)[(\tilde{A}+\kappa)y_{_{1}}]-2\eta^2\\
            \end{matrix}
            $
             \\
\hline
            $\widetilde{H^{(\kappa,\eta,\tilde{A})}_{_{4}}.VI}$&
            $\tilde{A}_{_{6}}$
            &
            $\gamma_{_{6}} C $
            &
            $
            \begin{matrix}
            \vspace{2mm}
            \tilde{A}=\eta=0,\\
            \xi_{_{6}}=-\frac{y_{_{1}} c_{_{1}}}{\tilde{x}_{_{2}}}(\frac{1}{2}+\kappa)+c_{_{2}}(\tilde{x}_{_{2}}),\\
            \gamma_{_{6}}=\frac{c_{_{1}}}{\tilde{x}_{_{2}}}\\
            \end{matrix}
            $
             \\
\hline
            $\widetilde{H^{(\kappa,\eta,\tilde{A})}_{_{4}}.VII}$&
            $\begin{pmatrix}
            a_{_{11}}(\tilde{x}_{_{1}}) & a_{_{12}}(\tilde{x}_{_{1}}) & a_{_{13}}(\tilde{x}_{_{1}}) & a_{_{14}}(\tilde{x}_{_{1}})\\
            \xi_{_{7}}                  & \xi_{_{7}}                  & \xi_{_{7}}                  & \xi_{_{7}}                 \\
            \gamma_{_{7}}               & \gamma_{_{7}}               & \gamma_{_{7}}               & \gamma_{_{7}}              \\
            a_{_{41}}(y_{_{2}})         & a_{_{42}}(y_{_{2}})         & a_{_{43}}(y_{_{2}})         & a_{_{44}}(y_{_{2}})        \\
            \end{pmatrix} $
            &
            $\lambda_{_{7}} C $
            &
            $
            \begin{matrix}
            \xi_{_{7}}=-\frac{\tilde{A}~y_{_{2}} c_{_{1}}}{\tilde{x}_{_{2}}^2}+c_{_{2}}(\tilde{x}_{_{2}}),\\
            \gamma_{_{7}}=-\frac{(1+2\kappa)c_{_{1}}}{2~\tilde{x}_{_{2}}}+c_{_{3}}(y_{_{1}}),\\
            \lambda_{_{7}}=\frac{c_{_{1}}}{\tilde{x}_{_{2}}},~\eta=0,
            \end{matrix}
            $
             \\
\hline
            $\widetilde{H^{(\kappa,\eta)}_{_{4}}.VIII}$&
            $\begin{pmatrix}
            a_{_{11}}(\tilde{x}_{_{1}}) & a_{_{12}}(\tilde{x}_{_{1}}) & a_{_{13}}(\tilde{x}_{_{1}}) & a_{_{14}}(\tilde{x}_{_{1}})\\
            a_{_{21}}(\tilde{x}_{_{2}}) & a_{_{22}}(\tilde{x}_{_{2}}) & a_{_{23}}(\tilde{x}_{_{2}}) & a_{_{24}}(\tilde{x}_{_{2}})\\
            \xi_{_{8}}                  & \xi_{_{8}}                  & \xi_{_{8}}                  & \xi_{_{8}}                 \\
            a_{_{41}}(y_{_{2}})         & a_{_{42}}(y_{_{2}})         & a_{_{43}}(y_{_{2}})         & a_{_{44}}(y_{_{2}})        \\
            \end{pmatrix} $
            &
            $\gamma_{_{8}} C $
            &
            $
            \begin{matrix}
            \eta=0,\\
            \xi_{_{8}} =\frac{c_{_{1}}(2\kappa-1)}{2(\tilde{x}_{_{2}}+\kappa y_{_{1}})}+c_{_{2}}(y_{_{1}}),\\
            \gamma_{_{8}} =\frac{c_{_{1}}}{\tilde{x}_{_{2}}+\kappa y_{_{1}}}\\
            \end{matrix}
            $
             \\
\hline
            $\widetilde{H^{(\kappa,\tilde{A})}_{_{4}}.X}$&
            $\tilde{A}_{_{10}} $
            &
            $\gamma_{_{10}} C $
            &
            $
            \begin{matrix}
            \xi_{_{10}}=\frac{-c_{_{1}}(-1+2(\tilde{A}-\kappa))}{2(\kappa-\tilde{A})(\tilde{x}_{_{2}}+(\kappa-\tilde{A})y_{_{1}})},\\
            \gamma_{_{10}}=\frac{c_{_{1}}}{\tilde{x}_{_{2}}+\kappa~y_{_{1}}-\tilde{A}y_{_{1}}}\\
            \end{matrix}
            $
           \\
\hline
\hline
\end{tabular}}}
\end{center}
\vspace{-2mm}
{~~\tiny $\tilde{A}_i=\begin{pmatrix}
            a_{_{11}}(\tilde{x}_{_{1}}) & a_{_{12}}(\tilde{x}_{_{1}}) & a_{_{13}}(\tilde{x}_{_{1}}) & a_{_{14}}(\tilde{x}_{_{1}})\\
            \xi_{_{i}}                  & \xi_{_{i}}                  & \xi_{_{i}}                  & \xi_{_{i}}                 \\
            a_{_{31}}(y_{_{1}})         & a_{_{32}}(y_{_{1}})         & a_{_{33}}(y_{_{1}})         & a_{_{34}}(y_{_{1}})        \\
            a_{_{41}}(y_{_{2}})         & a_{_{42}}(y_{_{2}})         & a_{_{43}}(y_{_{2}})         & a_{_{44}}(y_{_{2}})        \\
            \end{pmatrix} $,~$i=2, 4, 5, 6, 10$,~~~~~
            $C=\begin{pmatrix}
            0 & 0 & 0 & 0\\
            0 & 0 & 0 & 0\\
            1 & 1 & 1 & 1\\
            0 & 0 & 0 & 0\\
            \end{pmatrix} $.}
\\
\\
Finally, inserting \eqref{4.16} and \eqref{4.17} into equations \eqref{4.14} and \eqref{4.15} and then using
the fact that ${\tilde R}_{_{Ma}} = {\delta}_{_{Ma}}$, one can find the matrices ${\tilde A^{^a}}_{_{~b}}({\tilde x})$ and
${\tilde B^{^a}}_{_{~b}}({\tilde x})$. The result is
\begin{eqnarray}\nonumber
	{\tilde A^{^a}}_{_{~b}}({\tilde x}) = \left( \begin{array}{cccc}
		$$a_{_{11}}(\tilde{x}_{_{1}})$$  & $$a_{_{12}}(\tilde{x}_{_{1}})$$ &  $$a_{_{13}}(\tilde{x}_{_{1}})$$  & $$a_{_{14}}(\tilde{x}_{_{1}})$$ \\
		$$\xi$$                          & $$\xi$$                         &  $$\xi$$                          & $$\xi$$\\
		$$a_{_{31}}(y_{_{1}})$$          & $$a_{_{32}}(y_{_{1}})$$         &  $$a_{_{33}}(y_{_{1}})$$          & $$a_{_{34}}(y_{_{1}})$$ \\
		$$a_{_{41}}(y_{_{2}})$$          & $$a_{_{42}}(y_{_{2}})$$         &  $$a_{_{43}}(y_{_{2}})$$          & $$a_{_{44}}(y_{_{2}})$$ \\
	\end{array} \right),
\end{eqnarray}
\begin{eqnarray}\label{4.18}
	{\tilde B^{^a}}_{_{~b}}({\tilde x}) = \frac{1}{\tilde{x}_{_{2}}+\kappa y_{_{1}}}\left( \begin{array}{cccc}
		$$0$$                    & $$0$$                    &  $$0$$                    & $$0$$ \\
		$$c_{_{1}}$$             & $$c_{_{1}}$$             &  $$c_{_{1}}$$             & $$c_{_{1}}$$\\
		$$-c_{_{1}}+a_{_{0}}$$   & $$-c_{_{1}}+a_{_{0}}$$   &  $$-c_{_{1}}+a_{_{0}}$$   & $$-c_{_{1}}+a_{_{0}}$$ \\
		$$0$$                    & $$0$$                    &  $$0$$                    & $$0$$ \\
	\end{array} \right),
\end{eqnarray}
where $\xi=\frac{-3c_{_{1}}(1+\kappa)-a_{_{0}}(1+2~\kappa)}{2\kappa(\tilde{x}_{_{2}}+\kappa~y_{_{1}})}$.
Similarly, we obtain  the solutions of the equations \eqref{4.14} and \eqref{4.15} to all dual backgrounds of Table 4.
In this manner, we show that all dual models (except for the $\widetilde{H^{(\kappa,\eta,\tilde{A})}_{_{4}}.IX}$ model)
are integrable; the results are summarized in Table 5.

\vspace{3mm}
\begin{center}
\small {{{\bf Table 6.}~Non-Abelian T-dual backgrounds as solutions of the GSEs}}
{\scriptsize
\renewcommand{\arraystretch}{1.2}{
\begin{tabular}{llll} \hline \hline
Symbol &  Vector field $\tilde{I}$  & One-form $\tilde{Z}$ & Comments \\
\hline
\vspace{-2.5mm}\\		
                                    & $(\alpha_{_{2}}\rho-\alpha_{_{4}}\tilde{x}_{_{2}}+\alpha_{_{6}}) \partial{\tilde{x}_{_{1}}}$
                                    & $\Big[\frac{y_{_{1}}\alpha_{_{2}}+\alpha_{_{3}}}{\tilde{x}_{_{2}}+y_{_{1}}}\Big]d\tilde{x}_{_{1}}$   &  $\kappa=1$\\
                                    & $+(\alpha_{_{2}}\tilde{x}_{_{2}}-\alpha_{_{3}})\partial{\tilde{x}_{_{2}}}$
                                    & $+\frac{1}{\alpha_{_{2}}(\tilde{x}_{_{2}}+y_{_{1}})^2}\Big\{(\tilde{x}_{_{2}}+y_{_{1}})$     &  \\
                                    & $+(\alpha_{_{2}}y_{_{1}}+\alpha_{_{3}})\partial{y_{_{1}}}$
                                    & $\times(y_{_{1}}\alpha_{_{2}}\alpha_{_{4}}+2\alpha_{_{3}}\alpha_{_{4}}-\alpha_{_{2}}\alpha_{_{5}})$ &  \\
$\widetilde{H^{(\kappa)}_{_{4}}.I}$ & $+(-\alpha_{_{2}}\rho+\alpha_{_{4}}y_{_{1}}+\alpha_{_{5}}) \partial{y_{_{2}}}$
                                    & $+\alpha_{_{2}}(\tilde{x}_{_{2}}\alpha_{_{2}}-\alpha_{_{3}})\rho\Big\}d\tilde{x}_{_{2}}$ &  \\
                                    & & $-\frac{1}{\alpha_{_{2}}(\tilde{x}_{_{2}}+y_{_{1}})^2}\Big\{(\tilde{x}_{_{2}}+y_{_{1}})$ &\\
                                    & & $\times(\tilde{x}_{_{2}}\alpha_{_{2}}\alpha_{_{4}}-2\alpha_{_{3}}\alpha_{_{4}}+\alpha_{_{2}}\alpha_{_{6}})$ &\\
                                    & & $+\alpha_{_{2}}(y_{_{1}}\alpha_{_{2}}+\alpha_{_{3}})\rho\Big\}dy_{_{1}}$ &\\
                                    & & $+\Big[\frac{\tilde{x}_{_{2}}\alpha_{_{2}}-\alpha_{_{3}}}{\tilde{x}_{_{2}}+y_{_{1}}}\Big]dy_{_{2}}$ &\\
\vspace{-2.5mm}\\
\hline
\vspace{-2.5mm}\\
                                          & $[\alpha_{_{2}}(\rho-2\eta^2)-\alpha_{_{4}}\tilde{x}_{_{2}}+\alpha_{_{6}}] \partial{\tilde{x}_{_{1}}}$
                                          & $\Big[\frac{y_{_{1}}\alpha_{_{2}}+\alpha_{_{3}}}{\tilde{x}_{_{2}}+y_{_{1}}}\Big]d\tilde{x}_{_{1}}$            &  $\kappa=1$\\
                                          & $+(\alpha_{_{2}}\tilde{x}_{_{2}}-\alpha_{_{3}})\partial{\tilde{x}_{_{2}}}$
                                          & $+\frac{1}{\alpha_{_{2}}(\tilde{x}_{_{2}}+y_{_{1}})^2}\Big\{y_{_{1}}^2\alpha_{_{2}}\alpha_{_{4}}$     & \\
                                          & $+(\alpha_{_{2}}y_{_{1}}+\alpha_{_{3}})\partial{y_{_{1}}}$
                                          & $+2y_{_{1}}\alpha_{_{3}}\alpha_{_{4}}+\alpha_{_{2}}\alpha_{_{3}}(2\eta^2-\rho)$ &  \\
$\widetilde{H^{(\kappa,\eta)}_{_{4}}.II}$ & $+[\alpha_{_{2}}(\rho-2\eta^2)+\alpha_{_{4}}y_{_{1}}+\alpha_{_{5}}] \partial{y_{_{2}}}$
                                          & $+y_{_{1}}\alpha_{_{2}}(-\alpha_{_{5}}+4\alpha_{_{2}}\eta^2-2\alpha_{_{2}}\rho)$ &  \\
                                          & & $+\tilde{x}_{_{2}}(y_{_{1}}\alpha_{_{2}}\alpha_{_{4}}+2\alpha_{_{3}}\alpha_{_{4}})$ &\\
                                          & & $+\tilde{x}_{_{2}}\alpha_{_{2}}(-\alpha_{_{5}}+\alpha_{_{2}}(2\eta^2-\rho))\Big\}d\tilde{x}_{_{2}}$ &\\
                                          & & $+\frac{1}{\alpha_{_{2}}(\tilde{x}_{_{2}}+y_{_{1}})^2}\Big\{-\tilde{x}_{_{2}}^2\alpha_{_{2}}\alpha_{_{4}}$ &\\
                                          & & $+2y_{_{1}}\alpha_{_{3}}\alpha_{_{4}}+\alpha_{_{2}}\alpha_{_{3}}(2\eta^2-\rho)$ &  \\
                                          & & $-\tilde{x}_{_{2}}(y_{_{1}}\alpha_{_{2}}\alpha_{_{4}}-2\alpha_{_{3}}\alpha_{_{4}}+\alpha_{_{2}}\alpha_{_{6}})$ &\\
                                          & & $+y_{_{1}}\alpha_{_{2}}(-\alpha_{_{6}}+\alpha_{_{2}}(2\eta^2-\rho))\Big\}dy_{_{1}}$ &\\
                                          & & $+\Big[\frac{\tilde{x}_{_{2}}\alpha_{_{2}}-\alpha_{_{3}}}{\tilde{x}_{_{2}}+y_{_{1}}}\Big]dy_{_{2}}$ & \\
\vspace{-2.5mm}\\
\hline
\vspace{-2.5mm}\\
                                                     & $\alpha_{_{1}} \partial{\tilde{x}_{_{1}}}+\alpha_{_{3}} \partial{y_{_{1}}}+\alpha_{_{4}} \partial{y_{_{2}}}$
                                                     &
                                                     $\frac{1}{\tilde{x}_{_{2}}^2-\eta^2\rho^2}\Big[\big(\tilde{x}_{_{2}}\alpha_{_{3}}-\alpha_{_{4}}\eta^2\rho\big)
                                                     d\tilde{x}_{_{1}}$   & $k=1,$ \\
$\widetilde{H^{(\kappa,\eta,\tilde{A})}_{_{4}}.III}$ & & $-\big(\tilde{x}_{_{2}}(1-\alpha_{_{1}})+\alpha_{_{3}}\rho\big)d\tilde{x}_{_{2}}$ & $\tilde{A}=0$ \\
                                                     & & $+\tilde{x}_{_{2}}(-\alpha_{_{1}}+\alpha_{_{4}})dy_{_{1}}$               & \\
                                                     & & $+\big(-\tilde{x}_{_{2}}\alpha_{_{3}}+\alpha_{_{1}}\eta^2\rho\big)dy_{_{2}}\Big]$       & \\
\vspace{-2.5mm}\\
\hline
\vspace{-2.5mm}\\
$\widetilde{H^{(\kappa,\eta,\tilde{A})}_{_{4}}.IV}$ & $\frac{1}{\eta^2-1}(\alpha_{_{5}}\tilde{x}_{_{2}}+\alpha_{_{6}})\partial{\tilde{x}_{_{1}}}$
                                                    &$\frac{1}{(\tilde{x}_{_{2}}+y_{_{1}})(\eta^2-1)}\Big[\big(-y_{_{1}}\alpha_{_{5}}$ &
                                                    $\tilde{A}=-1+\kappa(1-\eta^2)+2\eta^2$ \\
                                                    & $+(\alpha_{_{5}}y_{_{1}}-\alpha_{_{7}})\partial{y_{_{2}}}$
                                                    & $+\alpha_{_{6}}+c_{_{1}}(\eta^2-1)\big) d\tilde{x}_{_{2}}$ & \\
                                                    & & $+\big({\tilde{x}_{_{2}}\alpha_{_{5}}+\alpha_{_{7}}+c_{_{1}}(\eta^2-1)}\big) dy_{_{1}}\Big]$ &\\
\vspace{-2.5mm}\\
\hline
\hline
\end{tabular}}}
\end{center}
\begin{center}
\small {{{\bf Table 6.}~Continued.}}
{\scriptsize
\renewcommand{\arraystretch}{1.2}{
\begin{tabular}{llll} \hline \hline
Symbol &  Vector field $\tilde{I}$  & One-form $\tilde{Z}$ & Comments \\
\hline
\vspace{-2.9mm}\\
                                                   & $\frac{1}{\eta^2-1}(\alpha_{_{5}}\partial{\tilde{x}_{_{1}}}+\alpha_{_{6}}\partial{y_{_{2}}}) $
                                                   & $\Big[-\frac{1}{\omega_{_{1}}}(y_{_{1}}\alpha_{_{2}}+\alpha_{_{3}})(\eta^2-1)\Big]d\tilde{x}_{_{1}}$ & $k=1,$ \\
                                                   & $+\Big[\frac{\alpha_{_{2}}(\tilde{x}_{_{2}}(\eta^2-1)-2\eta^2)}{\eta^2-1}\Big]\partial{\tilde{x}_{_{2}}}$
                                                   & $+\Big\{\frac{1}{\omega_{_{1}}^2}\Big[\alpha_{_{6}}\omega_{_{1}}$& $\tilde{A}=1-\kappa,$\\
$\widetilde{H^{(\kappa,\eta,\tilde{A})}_{_{4}}.V}$ & $+(\alpha_{_{2}}y_{_{1}}+\alpha_{_{3}})\partial{y_{_{1}}}$
                                                   & $+(y_{_{1}}\alpha_{_{2}}+\alpha_{_{3}})(-1+\eta^2)^2\rho\Big]\Big\}d\tilde{x}_{_{2}}$ & $\omega_{_{1}}=(\tilde{x}_{_{2}}+y_{_{1}})(1-\eta^2)$\\
                                                   & & $+\Big\{\frac{1}{\omega_{_{1}}^2}\Big[\alpha_{_{5}}\omega_{_{1}}-\rho(-1+\eta^2)$ & $~~+2\eta^2$\\
                                                   & & $\times \Big(\alpha_{_{3}}-(2\alpha_{_{2}}+\alpha_{_{3}})\eta^2$ &\\
                                                   & & $+\tilde{x}_{_{2}}\alpha_{_{2}}(\eta^2-1)\Big)\Big]\Big\}dy_{_{1}}$ & \\
                                                   & & $+\Big[\alpha_{_{2}}-\frac{(y_{_{1}}\alpha_{_{2}}+\alpha_{_{3}})(1-\eta^2)}{\omega_{_{1}}}\Big]dy_{_{2}}$ &\\
\vspace{-2.9mm}\\
\hline
\vspace{-2.9mm}\\
                                                    & $\alpha_{_{1}}\partial{\tilde{x}_{_{1}}}+\alpha_{_{3}} \partial{y_{_{1}}}$
                                                    &$\Big[\frac{\alpha_{_{3}}(1-\eta^2)}{\omega_{_{2}}}\Big]d\tilde{x}_{_{1}}$       &$k=1,$ \\
                                                    & $~~~+\alpha_{_{4}} \partial{y_{_{2}}}$
                                                    & $+\Big\{-\frac{1}{\omega_{_{2}}^2}\Big[\tilde{x}_{_{2}}(\eta^2-1)$ & $\tilde{A}=0,$\\
$\widetilde{H^{(\kappa,\eta,\tilde{A})}_{_{4}}.VI}$ & & $\times [\alpha_{_{1}}(\eta^2-1)+c1]$ & $\omega_{_{2}}=\tilde{x}_{_{2}}(1-\eta^2)+\rho\eta^2$\\
                                                    & & $-\rho\Big(\alpha_{_{3}}(-1+\eta^2)^2$ & \\
                                                    & & $+\eta^2\Big(-1+\eta^2+\alpha_{_{4}}(-1+\eta^2)+c_{_{1}}\Big)\Big)\Big]\Big\}d\tilde{x}_{_{2}}$ & \\
                                                    & & $+\Big[\frac{(\alpha_{_{1}}-\alpha_{_{4}})(\eta^2-1)}{\omega_{_{2}}}\Big]dy_{_{1}}$ &\\
                                                    & & $-\Big[\frac{\alpha_{_{3}}(1-\eta^2)}{\omega_{_{2}}}\Big]dy_{_{2}}$ &\\
\vspace{-2.9mm}\\
\hline
\vspace{-2.9mm}\\
$\widetilde{H^{(\kappa,\eta,\tilde{A})}_{_{4}}.VII}$ & $\alpha_{_{1}}\partial{\tilde{x}_{_{1}}}+\alpha_{_{4}} \partial{y_{_{2}}}$
                                                     & $\frac{1}{\tilde{x}^2(1+\eta^2)^2-\rho^2\eta^2}\Big[-\big(\alpha_{_{4}}\rho\eta^2(1+\eta^2)\big) d\tilde{x}_{_{1}}$ & $\tilde{A}=0,$ \\
                                                     & & $+\big({\tilde{x}_{_{2}}(\alpha_{_{1}}-1)(1+\eta^2)^2}\big)d\tilde{x}_{_{2}}$ & $\kappa=1$\\
                                                     & & $-\big({\tilde{x}_{_{2}}(\alpha_{_{1}}-\alpha_{_{4}})(1+\eta^2)^2}\big)dy_{_{1}}$ &\\
                                                     & & $+\big({\alpha_{_{1}}\eta^2(1+\eta^2)\rho}\big) dy_{_{2}}\Big]$ &\\
\vspace{-2.9mm}\\
\hline
\vspace{-2mm}\\
                                            & $\alpha_{_{1}}\partial{\tilde{x}_{_{1}}}+\alpha_{_{4}} \partial{y_{_{2}}}$
                                            & $\Big\{\frac{1}{\omega_{_{3}}}\Big[\tilde{x}_{_{2}}(-1+2c_{_{1}}+2\alpha_{_{1}})$ & $k=1-\eta^2,$ \\
                                            & & $+y_{_{1}}\Big(-1+2\alpha_{_{1}}(\eta^2-1)$ & $\lambda=\tilde{x}_{_{2}}+y_{_{1}},$\\
$\widetilde{H^{(\kappa,\eta)}_{_{4}}.VIII}$ & & $+2\alpha_{_{4}}\eta^2(\eta^2-1)+c_{_{1}}(2-4\eta^2)\Big)\Big]\Big\}d\tilde{x}_{_{2}}$ & $\omega_{_{3}}=2\lambda(\lambda-2y_{_{1}}\eta^2)$\\
                                            & & $+\Big\{\frac{1}{\omega_{_{3}}}\Big[\tilde{x}_{_{2}}\Big(-1+2c_{_{1}}+2\eta^2-2\alpha_{_{4}}(\eta^2-1)\Big)$ & \\
                                            & & $+y_{_{1}}\Big(-1+2\eta^2-2\alpha_{_{1}}\eta^2$ & \\
                                            & & $+c_{_{1}}(2-4\eta^2)+4\alpha_{_{4}}(\eta^2-1)^2\Big)\Big]\Big\}dy_{_{1}}$ &\\
\vspace{-2.9mm}\\
\hline
\vspace{-2.9mm}\\
                                              & $[-\alpha_{_{2}}\rho+\alpha_{_{4}}\tilde{x}_{_{2}}-\alpha_{_{6}}] \partial{\tilde{x}_{_{1}}}$
                                              & $\Big[-\frac{y_{_{1}}\alpha_{_{2}}+\alpha_{_{3}}}{\tilde{x}_{_{2}}-y_{_{1}}}\Big]d\tilde{x}_{_{1}}$            &  $\tilde{A}=1+\kappa$\\
                                              & $+(\alpha_{_{2}}\tilde{x}_{_{2}}+\alpha_{_{3}})\partial{\tilde{x}_{_{2}}}$
                                              & $+\frac{1}{\alpha_{_{2}}(\tilde{x}_{_{2}}-y_{_{1}})^2}\Big\{-y_{_{1}}^2\alpha_{_{2}}\alpha_{_{4}}$     & \\
                                              & $+(\alpha_{_{2}}y_{_{1}}+\alpha_{_{3}})\partial{y_{_{1}}}$
                                              & $-2y_{_{1}}\alpha_{_{3}}\alpha_{_{4}}-\alpha_{_{2}}\alpha_{_{3}}\rho-2y_{_{1}}\alpha_{_{3}}\alpha_{_{4}}$ &  \\
$\widetilde{H^{(\kappa,\tilde{A})}_{_{4}}.X}$ & $-[\alpha_{_{2}}\rho+\alpha_{_{4}}y_{_{1}}+\alpha_{_{5}}] \partial{y_{_{2}}}$
                                              & $+\tilde{x}_{_{2}}(y_{_{1}}\alpha_{_{2}}\alpha_{_{4}}+2\alpha_{_{3}}\alpha_{_{4}})$ &  \\
                                              & & $-\tilde{x}_{_{2}}\alpha_{_{2}}(\alpha_{_{5}}+\alpha_{_{2}}\rho)\Big\}d\tilde{x}_{_{2}}$ &\\
                                              & & $+\frac{1}{\alpha_{_{2}}(\tilde{x}_{_{2}}-y_{_{1}})^2}\Big\{-\tilde{x}_{_{2}}^2\alpha_{_{2}}\alpha_{_{4}}$ &\\
                                              & & $+2y_{_{1}}\alpha_{_{3}}\alpha_{_{4}}+\alpha_{_{2}}\alpha_{_{3}}\rho$ &  \\
                                              & & $+\tilde{x}_{_{2}}(y_{_{1}}\alpha_{_{2}}\alpha_{_{4}}-2\alpha_{_{3}}\alpha_{_{4}}-\alpha_{_{2}}\alpha_{_{6}})$ &\\
                                              & & $+y_{_{1}}\alpha_{_{2}}(\alpha_{_{6}}+\alpha_{_{2}}\rho)\Big\}dy_{_{1}}$ &\\
                                              & & $-\Big[\frac{\tilde{x}_{_{2}}\alpha_{_{2}}+\alpha_{_{3}}}{\tilde{x}_{_{2}}-y_{_{1}}}\Big]dy_{_{2}}$ & \\
\hline
\hline
\end{tabular}}}
\end{center}

\section{\label{Sec.V} Non-Abelian T-dual spaces of YB deformed $H_{4}$ WZW models as solutions of the GSEs}
In this subsection, we investigate that the dual models of Table 4 are also solutions of the GSEs. To this end, we
apply the method explained in subsection \ref{Sec.II.1} for the dual backgrounds of Table 4.
In this manner, we show that all the dual models (except for the $\widetilde{H^{(\kappa,\eta,\tilde{A})}_{_{4}}.IX}$) are solutions of the GSEs.
The results including vector fields $\tilde{I}$ and one-forms $\tilde{Z}$ are summarized in Table 6.
Note that for all cases, the cosmological constant vanishes.

\subsection{\label{Sec.V.1} Investigating the triviality of solutions of the GSEs of both original and dual models}

In Ref. \cite{W}, it has been analyzed that under what conditions solutions of the GSEs can be trivial in the
sense that they solve also the standard supergravity equations. There, it has been argued that for this to happen the
vector field $I$ must satisfy the following conditions
\begin{eqnarray}\label{5.1}
I^{^{M}}~I_{_{M}}=0,~~~~~~~dI=i_{_{I}}H,~~~~~~~X=d\Phi-I.
\end{eqnarray}

\begin{center}
\small {{{\bf Table 7.}~Triviality of solutions of the GSEs of original models}}
{\scriptsize
\renewcommand{\arraystretch}{1.6}{
\begin{tabular}{llll} \hline \hline
Symbol &  Vector field I  & One-form Z  & Comments \\
\hline		
$H^{(\kappa)}_{_{4}}.I$    & $\frac{-c_{_{1}}}{3}\partial_{_{v}}$               & $\frac{c_{_{1}}}{3}dx$     &  $\kappa=1$\\
\hline
$H^{(\kappa,\eta)}_{_{4}}.II$    & $\frac{-c_{_{1}}}{3}\partial_{_{v}}$               & $\frac{c_{_{1}}}{3}dx$ &  $\kappa=1$\\
\hline
$H^{(\kappa,\eta,\tilde{A})}_{_{4}}.III$    & $\frac{1}{2}(c_{_{3}}-c_{_{1}})\partial_{_{v}}$    & $c_{_{3}}dx$  &  $\tilde{A}=0,\kappa=1$\\
\hline
$H^{(\kappa,\eta,\tilde{A})}_{_{4}}.IV$    & $\frac{1}{2}(c_{_{3}}-c_{_{1}})(\eta^2-1)\partial_{_{v}}$    & $c_{_{3}}dx$    &  $\tilde{A}=1+\kappa(1-\eta^2)$\\
\hline
$H^{(\kappa,\eta,\tilde{A})}_{_{4}}.V$    & $\frac{1}{2}(c_{_{3}}-c_{_{1}})\partial_{_{v}}$    & $c_{_{3}}dx$              &  $\tilde{A}=1-\kappa$\\
\hline
$H^{(\kappa,\eta,\tilde{A})}_{_{4}}.VI$    & $\frac{1}{2}(c_{_{3}}-c_{_{1}})\partial_{_{v}}$    & $c_{_{3}}dx$             &  $\tilde{A}=0,\kappa=1$\\
\hline
$H^{(\kappa,\eta,\tilde{A})}_{_{4}}.VII$    & $\frac{1}{2}(c_{_{3}}-c_{_{1}})\partial_{_{v}}$    & $c_{_{3}}dx$             &  $\tilde{A}=0,\kappa=1$\\
\hline
$H^{(\kappa,\eta)}_{_{4}}.VIII$  & $\frac{c_{_{1}}}{3(\eta^2-1)}\partial_{_{v}}$    & $\frac{c_{_{1}}}{3}dx$        &  $\kappa=1-\eta^2$\\
\hline
$H^{(\kappa,\eta,\tilde{A})}_{_{4,q}}.IX$          & $\frac{(c_{_{3}}-c_{_{1}})(-1+q^4\eta^2)}{2(\eta^2-1)}\partial_{_{v}}$
                                                   & $\Big[c_{_{3}}+\frac{\eta^2(-1+q^4)[2+(-3+q^4)\eta^2]}{2(-1+q^4\eta^2)^2}x \Big]dx$  &  $\kappa=\frac{S_{_{1}}-S_{_{2}}}{(-1+q^4\eta^2)^2}$\\
\hline
$H^{(\kappa,\tilde{A})}_{_{4}}.X$    & $\frac{-c_{_{1}}}{3}\partial_{_{v}}$    & $\frac{c_{_{1}}}{3}dx$         &  $\tilde{A}=1+\kappa$\\
\hline
\hline
\end{tabular}}}
\end{center}
{\vspace{-2mm}
\tiny
$~~~~~~~~~~~~~~S_{_{1}}=\tilde{A}q^2[1-(1+q^4)\eta^2+q^4\eta^4],~~S_{_{2}}=\sqrt{(q^4\eta^2 -1)^2[1-2q^4\eta^2-(2-4q^4+q^8)\eta^4]}.$
}
\\
\\
In this subsection we examine the triviality of GSEs solutions of both original and dual models of Tables 2 and 6, respectively.
First, by simplifying the above relations and by inserting $X_{_{M}}= I_{_{M}}+ Z_{_{M}}$ into the last equation of \eqref{5.1} one can obtain
\begin{eqnarray}\label{5.4}
G_{_{M N}}I^{^{M}}I^{^{N}}=0,\\\label{5.5}
I_{_{M}}=\frac{1}{2}(\partial_{_{M}}\Phi-Z_{_{M}}),\\\label{5.6}
\partial_{_{M}}I_{_{N}}=\frac{1}{2}I^{^{P}}H_{_{P M N}}.
\end{eqnarray}
In this manner, we have shown that all solutions of the GSEs for the original models of Table 2 are trivial; the results are summarized in Table 7.
Furthermore, we have examined that none of the dual solutions are trivial.
Note that the dilaton field for all original models except for the $H^{(\kappa,\eta,\tilde{A})}_{_{4,q}}.IX$ is given by $c_{_{1}}x+c_{_{2}}$, while for the $H^{(\kappa,\eta,\tilde{A})}_{_{4,q}}.IX$ we find that
$\Phi=\{\eta^2(q^4-1)[2+(-3+q^4)\eta^2]x^2\}/\{4(-1+q^4\eta^2)^2\}+c_{_{1}}x+c_{_{2}}$
for some constants $c_{_{1}}, c_{_{2}}$.
\section{\label{Sec.VI} Conclusions}
In this paper, by using the constructed backgrounds of YB deformations of $H_{_{4}}$ WZW model in \cite{Eghbali1}
and following the general method outlined in the previous work \cite{Eghbali3},
we have shown that all deformed backgrounds can be considered as solutions of the GSEs.
Additionally, we have constructed the non-Abelian dual models of those models by the Poisson-Lie T-duality approach in the presence of spectator fields,
and have shown that those dual models, except for the $\widetilde{H^{(\kappa,\eta,\tilde{A})}_{_{4}}.IX}$ model, are integrable and satisfy the GSEs.
Following Wulff's work \cite{W}, we have discussed the triviality of solutions of the GSEs of both original and dual models of Tables 2 and 6, and have shown that only
solutions of original models of Table 2 are trivial.
Our results suggest that we can somehow conclude that the solutions of GSEs indeed remain invariant under the non-Abelian T-duality.
The method applied in this paper to examine the non-Abelian T-duality of the YB deformed $H_{_{4}}$ WZW models, was first used in \cite{Eghbali2}.
It would be interesting to generalize this method to a Lie supergroup case to calculate the non-Abelian target space duals
of the YB deformed WZW models based on the Lie supergroups \cite{EPR}.
We intend to address this problem in the future.

\subsection*{Declaration of competing interest}

The authors declare that they have no known competing financial
interests or personal relationships that could have appeared to influence the work reported in this paper.

\subsection*{Acknowledgements}

This work has been supported by the research vice chancellor of Azarbaijan Shahid Madani University under research fund No. 1402/231.
\\
\\
{\bf ORCID iDs}
\\
Ali Eghbali ~  https://orcid.org/0000-0001-6076-2179
\\
Simin Ghasemi-Sorkhabi ~ https://orcid.org/0009-0008-8821-4180
\\
Adel Rezaei-Aghdam ~ https://orcid.org/0000-0003-4754-7911

\subsection*{Data availability statement}

No data was used for the research described in the article.


\end{document}